\documentclass{emulateapj}

\shortauthors{Constantin et al. }  

\shorttitle{AGN in Voids}

\begin{document}

\slugcomment{ApJ in press; submitted on May 11, 2007}

\title{Active Galactic Nuclei in Void Regions}

\author{Anca Constantin} \affil{Department of Physics, Drexel
University, Philadelphia, PA 19104}

\author{Fiona Hoyle} \affil{Department of Physics and Astronomy,
Widener University, Chester, PA 19013}
                  
\author{Michael S. Vogeley} \affil{Department of Physics, Drexel
University, Philadelphia, PA 19104}

\begin{abstract}

We present a comprehensive study of accretion activity in the most
underdense environments in the universe, the voids, based on the SDSS
DR2 data.  Based on investigations of multiple void regions, we show
that Active Galactic Nuclei (AGN) are definitely common in voids, but
that their occurrence rate and properties differ from those in walls.
AGN are more common in voids than in walls, but only among moderately
luminous and massive galaxies ($M_r < -20$, log $M_*/M_{\sun} <
10.5$), and this enhancement is more pronounced for the relatively
weak accreting systems (i.e., $L_{\rm [O III]} < 10^{39}$ erg
s$^{-1}$).  Void AGN hosted by moderately massive and luminous
galaxies are accreting at equal or lower rates than their wall
counterparts, show lower levels of obscuration than in walls, and
similarly aged stellar populations.  The very few void AGN in massive
bright hosts accrete more strongly, are more obscured, and are
associated with younger stellar emission than wall AGN.  These trends
suggest that the accretion strength is connected to the availability of
fuel supply, and that accretion and star-formation co-evolve and rely
on the same source of fuel.  Nearest neighbor statistics indicate that
the weak accretion activity (LINER-like) usually detected in massive
systems is not influenced by the local environment.  However, H {\sc
ii}s, Seyferts, and Transition objects are preferentially found among
more grouped small scale structures, indicating that their activity is
influenced by the rate at which galaxies interact with each other.
These trends support a potential H{\sc~ii}$\rightarrow$Seyfert/Transition
Object$\rightarrow$LINER evolutionary sequence that we show is apparent in many
properties of actively line-emitting galaxies, in both voids and walls.
The subtle differences between void and wall AGN might be explained by
a longer, less disturbed duty cycle of these systems in voids.

\end{abstract}


\section{\sc Introduction}



The regions that are apparently devoid of galaxies \citep{kir81} and
clusters \citep{ein80}, the voids, are arguably the best probes
of the effect of the environment and cosmology on galaxy formation and
evolution.  If, as suggested by the standard cosmological paradigm,
structure in the present-day universe formed through hierarchical
clustering, with small structures merging to form progressively larger
ones, galaxies in the currently most underdense regions must be
the least ``evolved'' ones, as they must have formed at later times
than those in the dense regions.  Therefore, void and cluster galaxies
must follow different evolutionary paths.  Disturbing processes like
stripping and harassment, that operate preferentially in crowded
environments, should occur rarely in voids.  Studies of the properties
of the void galaxies, in contrast to those in relatively crowded
regions, or walls, should provide some of the strongest constraints
for distinguishing the intrinsic properties, which characterize a
galaxy when it is first assembled, from properties that have been
externally induced, over the whole history the universe: the ``nature
versus nurture'' problem.


Statistically significant conclusions regarding the distinctness of
the void galaxies relative to those in denser regions, the ``walls''
hereafter, emerged only recently, with the advent of large surveys
such as SDSS and 2dF.  Such data, and in particular SDSS, offered for
the first time the possibility to find and analyse both
photometrically and spectroscopically, large samples of extremely low
density regions (i.e., $\delta\rho/\rho < -0.6$ measured on a scale of
$7 h^{-1}$ Mpc, Rojas et al. 2004, 2005), and allowed for accurate
estimates of the void galaxy luminosity and mass functions
\citep{hoy05, gol05}.  These studies show that void galaxies are
fainter, bluer, have surface brightness profiles more similar to those
of late-type systems, and that their specific star formation rates are
higher than those in denser regions.  The mass and luminosity
functions are found to be clearly shifted towards lower characteristic
mass and fainter magnitudes ($M^*$).  Moreover, the faint end slopes
of the wall and void luminosity functions are very similar which
suggests that voids are not dominated by an excess population of
low-luminosity galaxies.  Consistently, no significant
excess in the amount of dark matter is apparent.  This means that, 
although largely devoid of light, the most underdense regions conform
to a galaxy formation picture which is clearly not strongly biased.


All these peculiarities demonstrate that the cosmological evolution of
void systems is different from that of those living in environments of
average cosmic densities.  Given the tight correlations between the
mass of black holes ($M_{\rm BH}$) and the dispersions and the masses
of the galactic bulges within which they reside \citep{mag98, fer00,
geb00, mar03}, one would then expect that the growth of massive BHs in
galaxy centers (and therefore the accretion process within active
galactic nuclei, AGN), also differs among distinct environs.
Extension of environmental studies of AGN properties to extreme
regions like cosmic voids is thus crucial to understand the
co-evolution of galaxies and their central BHs.  Moreover, while there
is general agreement that the growth of black holes must be closely
related to galaxy assembly \citep{silk98, kau00, beg05}, there is no
consensus as to how exactly accretion and star formation are coupled.
The void galaxies could be, arguably, the best test-bed for
understanding whether these processes are synchronized, or precede one
another, and whether feedback from the actively growing BHs
facilitates star formation (e.g., by dynamically compressing gas
clouds through radio jets), or suppresses it (e.g., by blowing away
the gas).


To date, studies of the spectral properties of the void AGN remain
limited to individual voids, e.g., the Bootes void \citep{kir81},
permitting the identification of only a few AGN among only a few dozen
void galaxies \citep{cru02}.  Quite surprisingly, such investigations
find that the AGN fraction and their emission-line properties are
similar in voids and in their field counterparts.   Moreover,
their associated stellar populations appear to share similar
characteristics in the two extreme environs.  The
conclusions of these studies are based on small number statistics and
do not however exclude the hypothesis that the void emission-line
activity, whether originating in star-formation or accretion, could be
connected with, e.g., filaments within voids; such structures would
provide local environs similar to those in the field.  The present SDSS
samples of voids and void galaxies offer us for the first time the
possibility to test and observationally constrain such ideas.

It is important to note that previous investigations of the
environmental dependence of nuclear activity in the relatively nearby
universe do not reach the extreme spatial densities representative of
voids.  For example, in \citet{kau04}, the lowest density regions
include over 25\% of the galaxies, which is more than 3 times more
galaxies than the true void regions encompass.  Their conclusions are
interesting, and an extension of such an investigation at truly low
densities is clearly desirable.  In particular, it is important to
quantify the degree to which the finding that, at fixed stellar mass,
twice as many galaxies host strong-lined AGN in low-density regions
than in high, extends to cosmic voids.  Our work provides such an
analysis.

We employ in this work the most accurately classified samples of voids
identified within SDSS to date, that yield $\sim 10^3$ void galaxies.
Motivated by our recent study on the AGN clustering phenomena
\citep{con06a}, which shows that there are differences in the large
scale structure of active galaxies, and that their clustering
amplitude correlates with their strength or rate of accretion, and
possibly with the availability of fuel, we compare void and wall
active galaxies of different types as classified based on their
emission-line properties.  Through such a comparison we aim to
understand: 1) how the large and small-scale structures influence
accretion onto their central black holes, and 2) to what degree AGN
activity is triggered by interactions or mergers between galaxies.  To
answer these questions, we investigate the occurrence rate of
different types of spectrally defined AGN, and how their accretion
activity relates to their associated black hole mass, the mass and the
age of their associated stellar populations, host morphology,
brightness, and nearest-neighbor distance.

We organize the paper as follows.  In Section ~\ref{data} we present
the void and wall sample selection, and the spectral classification we
use in defining various types of actively emitting galaxies.  We
compare and discuss the AGN occurrence rate in voids and walls, both
globally and at fixed host properties in Section ~\ref{fractions}.  We
examine in Section ~\ref{voidaccretion} the accretion rates, the fuel
supply, and the properties of the associated star-formation in void
and wall actively line emitting systems, while in Section
~\ref{nnprop} we discuss potential differences in their small scale
environments.  We summarize our findings in ~\ref{discussion}, and
discuss the possible implications on the nature of the power sources
in the low luminosity AGN, the AGN--host connection, and current
models of galaxy formation.  In particular, we show empirical evidence
for a possible evolutionary sequence that links different types of strong
line-emitting galaxies defined based on their spectral
characteristics.  Throughout this work, unless otherwise noted, we
assume $\Omega_m = 0.3$, $\Omega_{\Lambda} = 0.7$, and $H_0 = 100h$ km
s$^{-1}$ Mpc$^{-1}$.

\section{\sc The Data} \label{data}

We employ for this study $10^3$ void galaxies identified from the
large-scale structure {\it sample10}, as described in the NYU
Value-Added Galaxy Catalog \citep{bla05}, which is a subset of the
SDSS Data Release 2 that \citet{str02} describe in detail.  This
sample covers nearly 2000 deg$^2$ and contains 155,126 galaxies.
Technical details about the photometric camera, photometric analysis
and photometric system employed by SDSS can be found in \citet{gun98},
\citet{lup02}, and \citet{fuk96} and \citet{smi02}, respectively.
\citet{hog02} describe the photometric monitor, \citet{pie03} present
the astrometric calibration, and \citet{eis01}, \citet{str02} and
\citet{bla03} discuss the selection of the galaxy spectroscopic
tiling.  An overview of the SDSS can be found in \citet{yor00}.

The following subsections summarize the method of identification of
void and wall galaxies, and their general properties.  We present them
in terms of emission-line activity and their classification in various
species that reflect various degree of contribution from nuclear
accretion and star formation.  Our spectral analysis uses measurements
of absorption and emission line fluxes and equivalent widths (EW)
drawn from a catalog built by the MPA/JHU
collaboration\footnote{publicly available at {\tt
http://www.mpa-garching.mpg.de/SDSS/} \citep{bri04}}.  In this
dataset, the line emission component is separated and subtracted from
the total galaxy spectrum based on fits of stellar population
synthesis templates ~\citep{tre04}.  To relate the central BH
accretion activity to the host properties, we also use in this
analysis stellar masses of galaxies and two stellar absorption-line
indices, the 4000\AA\ break strength and the H$\delta_A$ Balmer
absorption-line index, as calculated by \citet{kau03}.  A detailed
analysis of many of these properties, and their relation to the AGN
phenomenon in particular, is presented in \citet{kau04}.

\subsection{The Void Galaxy Sample}

Voids galaxies were identified using a nearest neighbor analysis to
find galaxies that reside in regions of density contrast with $\delta
\rho / \rho < -0.6$ as measured on a scale of 7$h^{-1}$Mpc.  Wall
galaxies are then those objects for which $\delta \rho / \rho \ge
-0.6$.  The void galaxy selection procedure is described in detail in
\citet{rojasphoto}.  Briefly, this method uses a volume-limited sample
( $z_{\rm min}=0.034$, and $z_{\rm max}=0.089$) of relatively bright
galaxies ($M_r < -19.5$) to define the density field that traces the
distribution of voids.  From the flux limited sample that extends to
this $z_{\rm max}$, the galaxies with fewer than three neighbors in
the volume limited sample, within a sphere of radius 7$h^{-1}$Mpc, are
flagged as belonging to voids while the rest of them are classified as
wall galaxies.  This procedure yields a sample of 1,010 void galaxies
and 12,732 wall galaxies.  Objects that lie close to the edge of the
survey have been discarded as it is impossible to accurately count the
neighbors if they are not observed.

Note that the density around void galaxies ($\rho_{vg}$) is higher
than the mean density of a void ($\bar{\rho}_{void}$) because galaxies
are clustered and the few void galaxies tend to lie close to the edges
of the voids.  Our choice of density contrast and nomenclature is
consistent with studies of voids in other three-dimensional samples
\citep{HV02,HV04} where individual void structures were identified
using an objective {\tt voidfinder} algorithm that measures void
sizes, average densities, and density profiles.  This procedure
uncovers voids with typical radii of 12.5$h^{-1}$Mpc that fill 40\% of
the Universe and have a mean density $\delta \rho / \rho < -0.9$.  The
average density around the few galaxies in voids is typically $\delta
\rho / \rho < -0.6$ when measured on a scale of 7$h^{-1}$Mpc (the
typical underdensity for voids is $\delta \rho / \rho \approx -0.78$).
Tests of such void-finding methods with cold dark matter simulations
and semi-analytic models \citep{ben03} indicate that such
identifications and the locations of true voids are accurate in the
distributions of both galaxies and mass.

\subsection{The Spectral Classification}

We identify and classify accretion sources and other types of active
systems in the void and wall galaxy samples based their emission line
properties.  Following, \citet{bpt81}, \citet{vei87}, and
\citet{con06a}, we employ a set of four line flux ratios to
distinguish between systems in which ionization is dominated by
accretion and/or starlight: [\ion{O}{3}]$\lambda$5007/H$\beta$,
[\ion{N}{2}]$\lambda$6583/H$\alpha$,
[\ion{S}{2}]$\lambda\lambda$6716,6731/H$\alpha$, and
[\ion{O}{1}]$\lambda$6300/H$\alpha$.  Thus, we select from the parent
sample of galaxies a subset of strong emission-line sources that show
significant emission in all six lines used in the type classification
(H$\alpha$, H$\beta$, [\ion{O}{3}],[\ion{N}{2}], [\ion{S}{2}], and
[\ion{O}{1}]), and a set of passive objects that show insignificant,
if any, line emission activity.  An emission feature is considered to
be significant if its line flux is positive and is measured with at
least 2$\sigma$ confidence.

\begin{figure}
\plotone{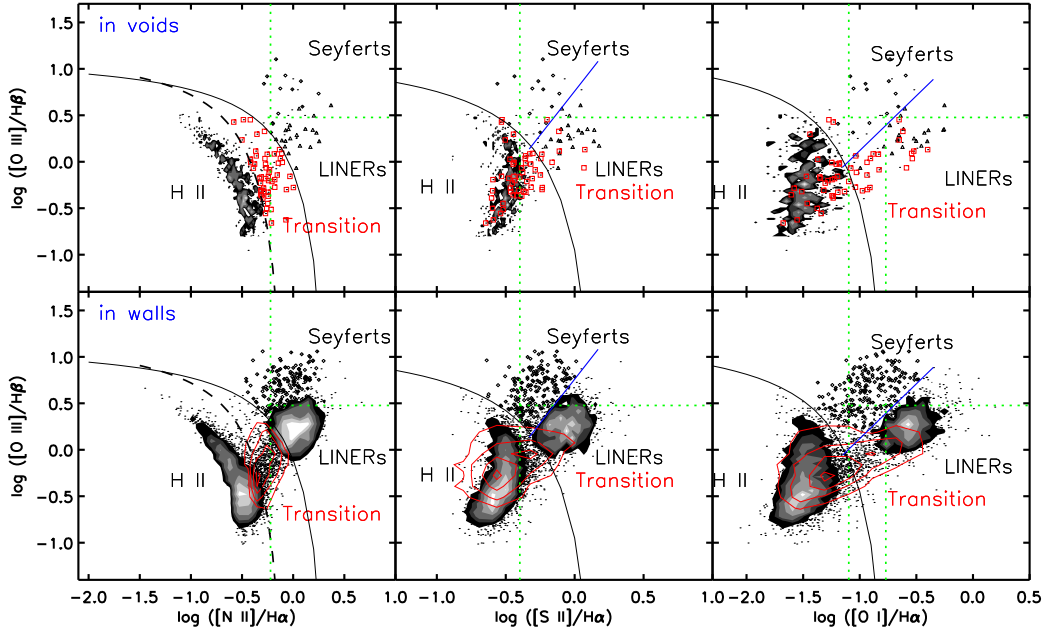}
\caption{ Diagnostic diagrams for emission-line galaxies in voids
  and walls, for objects with relatively high ($> 2$) signal-to-noise
  line flux measurements.  With the exception of void and wall
  Seyferts, and void LINERs and Transition objects, which are plotted
  individually, object types are shown in density contours
  corresponding to factors of $n$ of the total number of objects in
  each class, where $n = 0.1, 0.2, 0.3, 0.5, 0.7, 0.9$ starting from
  outermost contour.  The solid and dashed black curves illustrate the
  \citet{kew01} and \citet{kau03} separation lines, the blue diagonal
  lines illustrate the separation between Seyferts and LINERs by
  \citet{kew06}, while the green dotted lines indicate the
  \citet{ho97} classification criteria.
\label{bpt05}}
\end{figure}

To classify the actively line emitting objects we employ the criteria
proposed by \citet{kew06}.   Through their semi-empirical
nature, the \citet{kew06} separation lines match well with objects'
locations in three diagnostic diagrams which combine pairs of the six
lines mentioned above.  Figure ~\ref{bpt05} illustrates these definitions
for cases where the line fluxes are constrained to $> 2\sigma$
accuracy, separately in voids and walls.  The galaxies that show
strong star-formation activity, the H{\sc~ii} galaxies, are those
that lie below the ~\citet{kau03} curve (the dashed line) in the
[\ion{O}{3}]/H$\beta$ vs. [\ion{N}{2}]/H$\alpha$ diagram, and remain
confined to the left wing of the point distribution in the other two
diagrams as well.  Objects situated above this curve but below the
~\citet{kew01} theoretical ``maximum star-formation'' line (continuous
black line) are defined as Transition objects (Ts, or Composites).
Galaxies whose line flux ratios place them above the ~\citet{kew01}
curves are those where the AGN component is considered to be dominant.
The AGN are separated into Seyferts (Ss) and LINERs (Ls) by a diagonal
(continuous blue line) that represents the best fit to the location of
the minimum of the distributions in pairs of line ratios defining the
[\ion{O}{3}]/H$\beta$ vs. [\ion{S}{2}]/H$\alpha$, and
[\ion{O}{3}]/H$\beta$ vs. [\ion{O}{1}]/H$\alpha$ diagrams
\citep{kew06}.

Even though we consider these criteria as the best basis for
categorizing the line-emitters, we test our results with respect to
the \citet{ho97} definition, for the sake of comparison with previous
works.  Figure ~\ref{bpt05} shows the \citet{ho97} separation lines as
as (green) dotted lines.  There are some important differences between
these two classification methods, some of which are briefly discussed
by \citet{kew06}.  Although the LINER samples remain at least 90\%
consistent in these two different definitions, a significant number of
Seyferts and Transition objects change class.  The Kewley et
al. criteria classify quite a few Seyferts that would be defined by Ho
et al. either as H{\sc~ii}s or Ts based on their low ionization level,
gauged by their [\ion{O}{3}]/H$\beta$ ratio.  The number of objects
classified as Ts significantly increases with the \citet{kew06}
method.  However, many of these systems are clearly dominated by
starlight, even if their emission embodies a possible accretion
component.  \citet{con06a} show that a large fraction of Ts defined
based only on the [\ion{N}{2}]/H$\alpha$ diagram heavily populate the
H{\sc~ii} locus in the other two diagrams.  The [\ion{S}{2}]/H$\alpha$
and [\ion{O}{1}]/H$\alpha$ line flux ratios are more sensitive to the
type of ionization than [\ion{N}{2}]/H$\alpha$, and therefore better
discriminators between accretion and stellar dominated excitation
processes; hence, the AGN definitions based on the
[\ion{O}{3}]/H$\beta$ vs. [\ion{N}{2}]/H$\alpha$ diagram should be
regarded with caution.  For the void galaxy sample, however, the
separation between LINERs and Seyferts is practically independent of
the chosen classification method; the sample is very small in size and
the fraction of border-line objects is relatively low.

We note here that LINERs are not unambiguously considered to be AGN.
Mechanisms alternative to accretion onto a black hole, like
photoionization by hot, young stars \citep{fil92, shi92, bar00},
clusters of planetary nebula nuclei \citep{tan00}, or shocks
\citep{dop95}, have proved relatively successful in explaining the
optical spectra of these sources.  Particularly ambiguous in their
interpretation are the Transition sources.  It is not clear whether
these objects are a "composite," with a LINER/Seyfert nucleus
surrounded by star-forming regions, or not.  \citet{shi06} and
\citet{con07} show examples where this model does not work.  Thus, we
consider that among the narrow-line emitting objects only Seyferts are
bona-fide AGN, and advise more caution in generalizing the term AGN.

\section{\sc Incidence of AGN Activity} \label{fractions}

We present here the results of a comparative analysis of both
photometric and spectroscopic properties of the void and wall
galaxies.   We consider separately various types of emission-line 
activity, whether dominated by star-formation, accretion, or a mix of
them.  In this section in particular, we compare the rate of
occurrence/detection of different types of emission-line activity in
voids and walls, both globally and at fixed host properties.

\subsection{Global Frequencies} \label{fracglob}

We present in Table ~\ref{statistics} the percentages of strong
line-emitters (all 6 lines are detected) and subclasses of objects
relative to the whole void and wall galaxy samples.  We list these
numbers for definitions involving strictly high statistical
significance in the line flux measurements (greater than $2-\sigma$
level).  We find that strong line-emission activity is clearly more
common in voids than in walls.  This difference seems to be accounted
for by the difference in the rate of occurrence of H{\sc~ii}-type
emission.  Objects of other types of spectral activity show equal or
lower frequency in the most underdense regions relative to the crowded
ones.  That the fraction of H{\sc~ii} galaxies in the void regions
($32.8\%$) is significantly higher than in walls ($20.8\%$) is
consistent with previous spectroscopic studies of void galaxies
indicating that their specific star formation rates are higher than in
their wall counterparts \citep{rojasspectro}, and with the finding
that the H{\sc~ii}s are less clustered than other galaxies
\citep{con06a}.

The systems whose spectra indicate the presence of an accreting black
hole as an ionization source show a somewhat mixed behavior.  While
Seyferts seem to be equally represented in voids and walls ($1.5\%$),
their weaker counterparts, the LINERs, are present to a significantly
smaller fraction in voids ($2\%$) than in walls ($4.1\%$).  Although
the samples of objects involved in this comparison are small, with
only 20 LINERs in voids and 510 LINERs in walls, the difference is
significant at more than $3-\sigma$.  \citet{con06a} show that LINERs
cluster more strongly than Seyferts, so the weaker representation of
LINERs in the void regions is not surprising, and supports these
results.
The Transition objects show an intermediate behavior between LINERs
and Seyferts, being slightly less common in voids than in walls. These
trends are somewhat expected based on their clustering properties as
well, although the significance of this comparison is only at the
$1.5-\sigma$ level.

It is important to note that these trends are nearly independent of
the classification criteria.  The drop in the fraction of LINERs from
walls to voids is only slightly larger using the \citet{ho97}
classification, from $1.8\%$ (18 out of 996 void galaxies) to $4.4\%$
(552 out of 12153 wall galaxies), confirming thus that this type of
activity is not frequent in the lowest density regions.  Even for the
Transition objects, whose definition varies the most between
\citet{ho97} and \citet{kew06}, with the latter allowing for more
objects that are either classified as H{\sc~ii}s or remain
unclassified by the first method, the fractions show the same trend,
being higher in walls, and do not change by more than 5\%.  Thus, the
results are consistent with those based on the \citet{kew06}
definition, and show that AGN activity is generally less common in
voids than in walls.

\subsection{Frequencies at fixed Host Properties} \label{fracprop}

Because void and wall galaxies are characterized by quite different
distributions in their morphologies and luminosities
\citep{rojasphoto}, any possible variation in the AGN properties,
including the frequency of their occurrence, could be related to such
morphological differences.  To examine these issues for our particular
samples, we present in Table ~\ref{host} the median values for host
$r$-band absolute magnitudes ($M_r$) and host concentration indices
($C$\footnote{$C = R_{50}/R_{90}$, where $R_{50}$ and $R_{90}$ are the
radii from the center of a galaxy containing 50\% and 90\% of the
Petrosian flux measured in $r$-band.} as a proxy for the morphological
type) for void and wall line-emitting objects of different spectral
types.  Void hosts are clearly less luminous than their wall
counterparts, by $\approx 0.5$ mag, although their morphologies are
quite similar.  There is no evidence for more late type systems
(smaller $C$ values) hosting actively line-emitting galaxies in voids
than in walls.  We will thus emphasize here a comparison of various
emission properties at fixed host brightness.

\begin{figure}
\plottwo{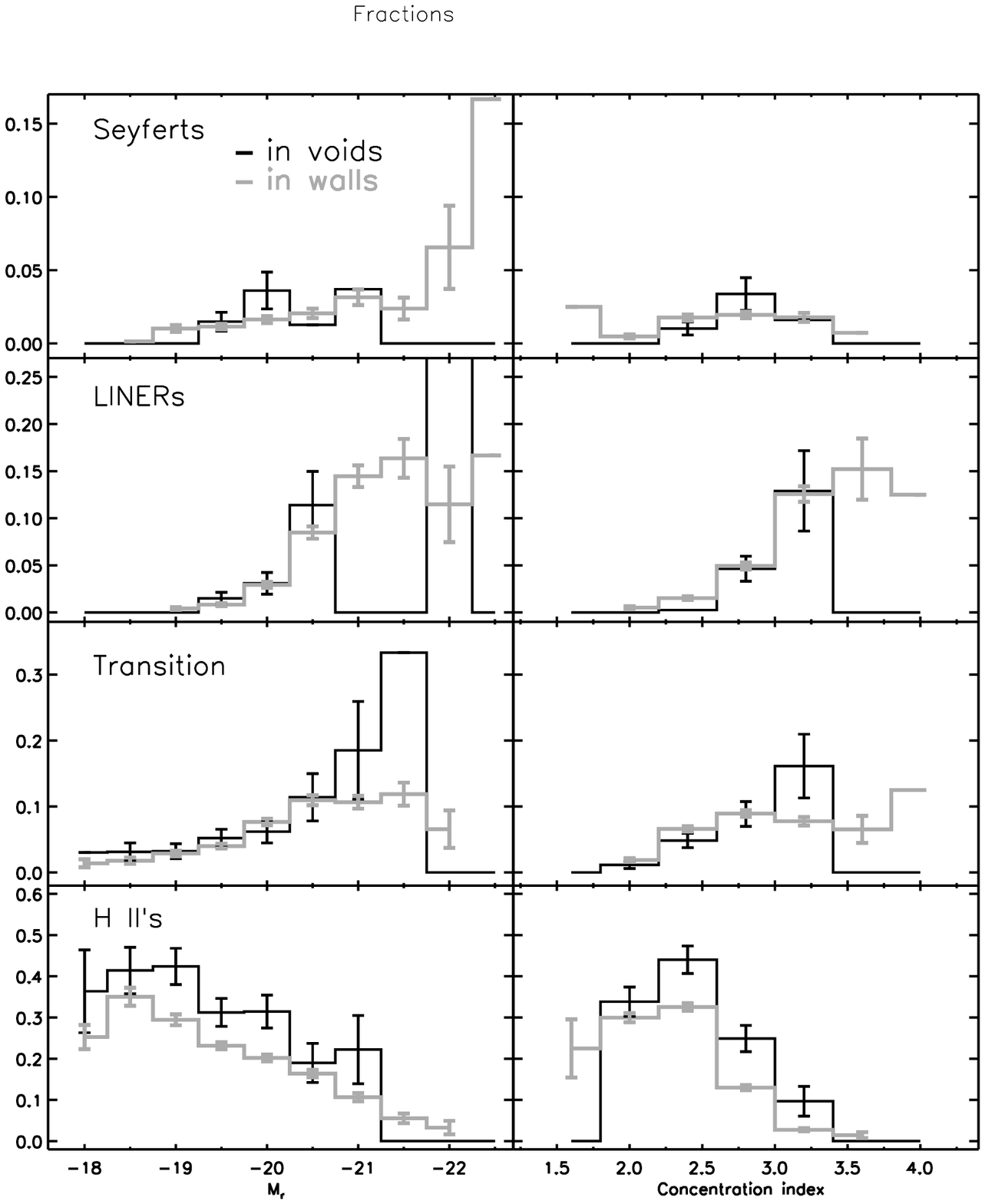}{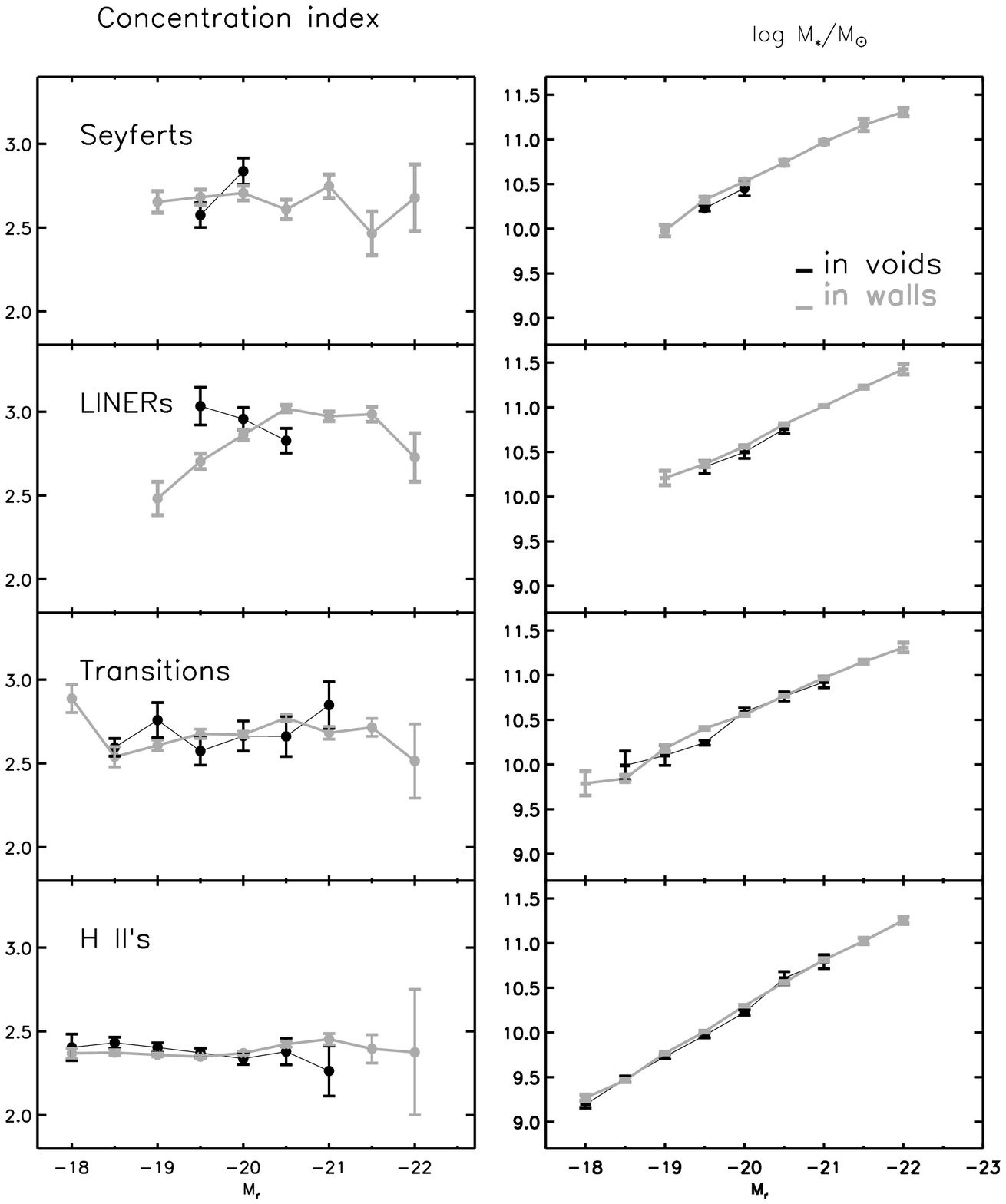}
\caption{{\it Left}: Fractions of void and wall galaxies that are
  spectroscopically classified as Seyferts, LINERs, Transition
  objects, and H{\sc~ii}s, as a function of their $r$-band absolute
  magnitude ($M_r$) and their concentration indices ($C$).  {\it
  Right}: Median values for $C$ and (dust corrected) stellar mass (log
  $M_*/M_{\sun}$) in bins of 0.5 mag of $M_r$, separately in voids
  (black) and walls (gray). Error bars are shown only for bins that
  include at least two objects.
\label{fracmag}}
\end{figure}

Figure ~\ref{fracmag} illustrates the fractions in which Seyferts, Ls,
Ts and H{\sc~ii}s are detected in systems characterized by different
$M_r$ and $C$ values.  We also show here how $C$ and the stellar mass
(log $M_*/M_{\sun}$) are distributed as a function of $M_r$ for all
these kinds of objects.  Note that the stellar mass and the intrinsic
brightness of galaxies correlate very well in both voids and walls.
Thus, our investigation of AGN fractions and other emission-line
properties at fixed brightness apply as well to the corresponding bin
in stellar mass.  Throughout the paper, we will interchangeably
characterize galaxies by stellar mass or absolute magnitude.  The
distribution of $C$ as a function of $M_r$ is generally flat given the
errors, and there is no significant difference in the $C$ values of
void and wall systems at any given brightness.  LINERs are an
exception: the ones in walls show earlier type hosts at larger
brightness while the ones in void exhibit the opposite trend, giving
rise to significant differences in $C$ between void and wall systems
of a given $M_r$.
 
When fractions are compared at fixed luminosity and fixed
concentration index some interesting trends stand out.  The most
striking feature is that the objects that show signs of AGN activity,
i.e., the Seyferts, Ls, and Ts, are clearly underrepresented among the
most luminous or massive ($M_r < -20$, log $M_*/M_{\sun} > 10.5$)
hosts in voids.  Among moderately bright, medium mass galaxies ($M_r
\approx -20.5$, log $M_*/M_{\sun} \approx 10.5$), Seyferts appear more
frequently in voids than in walls, while Ls and Ts are equally common
in these different environs.
For H{\sc~ii}s, their frequencies among void galaxies are clearly
higher than in wall galaxies, at almost all host brightnesses and
morphologies.  This result agrees well with previous findings of
higher specific star formation rates in void galaxies relative to
their wall counterparts \citep{rojasspectro}, with the general
tendency of actively star-forming systems to be more common in the
most underdense regions than in other environs, and in particular with
the fact that H{\sc~ii}s are less clustered than other galaxies
\citep{con06a}.

It is noteworthy that Seyferts appear to favor different kinds of
hosts in voids and walls.  In voids, they concentrate almost
exclusively in galaxies with $M_r \approx -20$, spanning not more than
an order of magnitude in their luminosities ($-19.5 \ga M_r \la
-20.5$).  In walls, Seyferts prefer generally bright hosts.  While the
fraction of Seyferts in luminous wall systems increases at least 3
times relative to that in fainter hosts, there are no detected void
Seyferts at $M_r \la -20.5$.  On the other hand, where void and wall
Seyfert galaxies overlap in brightness, Seyfert-like activity is more
frequent in underdense regions than in more crowded environments.
Finally, void Seyferts seem to prefer earlier type galaxies than those
living in walls (there are no void Seyferts in galaxies with $C \la
2.25$).

LINERs and Ts are peculiar as well; while their fractions among faint
and moderately bright hosts are quite similar in voids and walls, they
basically disappear among the bright void galaxies.  The spike at $M_r
\approx -22$ corresponds to the only void galaxy that exhibits this
level of brightness.  Thus, the large difference in the total fraction
of LINERs between voids and walls is explained solely by the strong
deficiency of such systems in luminous hosts in underdense regions.
A similar trend in the occurrence rate at fixed brightness is present
among Ts, however, they seem to show peculiarities as a function of in
their morphological type as well.  Void Ts hosted by earlier type
galaxies ($C \approx 3$) seem more frequent than the wall ones, but
they are similarly common in void and wall hosts of late-type
morphologies.  This is interesting given that void emission-line
systems are generally absent among early-type hosts.

There are only three actively line-emitting void galaxies at $M_r
\approx -21$.  One object is spectrally classified as a T, while the
other two remain unclassified because they do not comply
simultaneously to the Seyfert-LINER separation criteria in the
diagrams involving [\ion{S}{2}] and [\ion{O}{1}], and thus have
properties that are somewhat intermediate between a Seyfert and a
LINER.  Even if these two objects are both Seyferts or both Ls, there
is no statistically significant rise in the fractions of void AGN
corresponding to this brightness range.  The error bars of these
fractions remain rather large; with either 2 Ls or 2 Seyferts in this
absolute magnitude range, their fraction in voids would be $0.3\% \pm
0.3\%$.  Thus, among bright galaxies, the relative fraction of
accretion-dominated systems in voids and walls remains unconstrained.  
Moderately and less luminous void galaxies are however clearly more prone
to hosting AGN than their wall counterparts.

\subsection{Frequencies and the Strength of Accretion Activity}

That void and wall galaxies host accretion activity to a different
degree should not come as a surprise.  \citet{kau04} showed that at
fixed galaxy mass, the low-density regions host twice as many AGN as
the high density ones.  We find here, however, that such a trend
extends to voids $only$ for galaxies that are moderately bright and
moderately massive ($M_r \approx -20$, $10 <$ log $M_*/M_{\sun} < 10.5$).
Figure ~\ref{fracmag} shows that, while the fraction of galaxies
containing strong AGN increases with brightness, and therefore with
$M_*$, in both voids and walls, it is clear that there is no
statistically significant surplus of any type of AGN in the most
massive void galaxies relative to their wall counterparts.

\begin{figure}
\plottwo{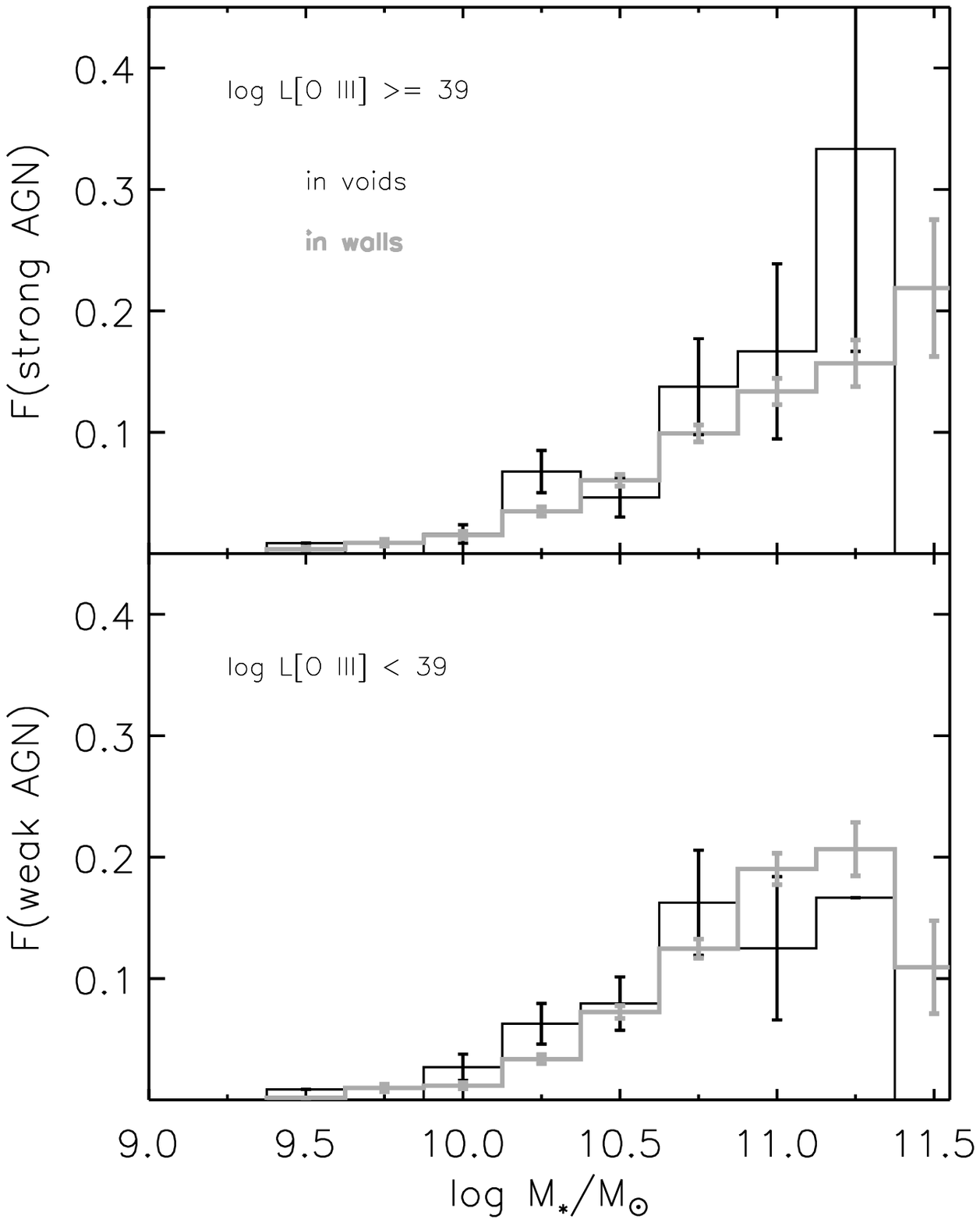}{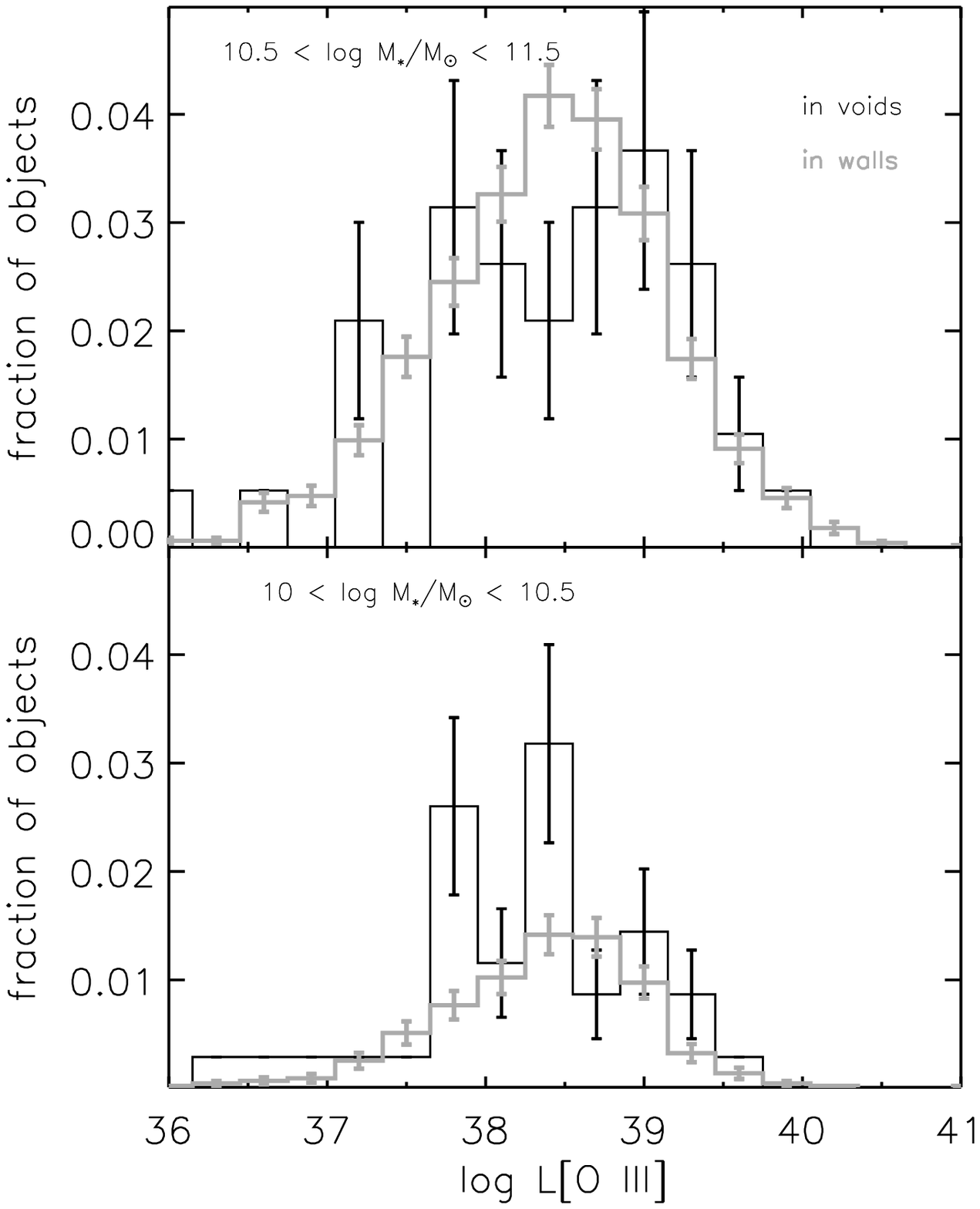}
\caption{ {\it Left}: Fractions of galaxies containing strong and weak
  AGN, with log $L_{\rm [O III]}$/(erg s$^{-1}$) higher and lower than
  39 respectively, plotted as a function of stellar mass, for voids
  (black) and walls (gray).  {\it Right}: Comparison of void and wall
  distributions in [\ion{O}{3}] line luminosities of objects with
  potential contribution from accretion (Ss + Ls + Ts), shown
  separately for two ranges in the host stellar mass.  The
  [\ion{O}{3}] luminosities have been corrected for internal dust
  attenuation as described in the text, and are expressed in erg
  s$^{-1}$.
\label{lumo3mu}}
\end{figure}

For a more direct comparison with \citet{kau04} results, we also
examine the fractions in which void and wall galaxies of a given
stellar mass host AGN.  Using the exact definition of AGN samples
employed by \citet{kau04}, we show these fractions in Figure
~\ref{lumo3mu}, separately for strong and weak [\ion{O}{3}] line
emitters.  It is readily apparent that AGN activity at all levels is
equally frequent among massive void and wall galaxies (log
$M_*/M_{\sun} > 10.5$), while at lower mass (log $M_*/M_{\sun} <
10.5$), both strong and weak AGN are generally more frequent in voids
than in walls.  The total fraction of massive (log $M_*/M_{\sun} >
10.5$) galaxies with strong lines ($L_{\rm [O III]} > 10^{39}$ erg
s$^{-1}$) is 9\% in walls and only 11\% in voids, which is far from
the 100\% difference found by \citet{kau04}.  On the other hand, for
the log $M_*/M_{\sun} < 10.5$ galaxies, the fraction of strong-lined
AGN in voids (2.8\%) is twice as high as that in walls (1.4\%).
Figure ~\ref{lumo3mu} also shows the distributions of [\ion{O}{3}]
luminosities in objects of high and medium stellar mass ranges,
separately for void and wall galaxies.  Among the massive galaxies,
only the low-accretion (log $L_{\rm [O III]}$/erg s$^{-1}$ $\approx
37$) systems appear marginally more frequent in voids than in walls,
while all levels of AGN activity seem more common in voids than in
walls among less massive hosts.

To summarize, the environmental dependence of the rate of occurrence
of the AGN phenomenon varies with the properties spanned by their
hosts.  Our comparative analysis conducted at fixed host properties
shows clearly that nuclear accretion is more frequent in voids than in
walls among moderately massive and luminous ($M_r \sim -20$, log
$M_*/M_{\sun} < 10.5$) hosts while it remains similarly common among
massive, luminous ($M_r \la -20$ and log $M_*/M_{\sun} > 10.5$) void
and wall galaxies.  These results show that accounting for host galaxy
properties, i.e., luminosity and mass, is essential for understanding
the environmental behavior of AGN activity; previous findings of
roughly constant fraction of galaxies hosting AGN across a wide range
of environments \citep{mil03} are justified because they are based on
investigations of bright galaxies ($M_r \leq -20$) only.

\subsection{Incidence of radio activity} \label{fracradio}

Scrutiny of potential differences in radio activity between wall and
void galaxies is of great interest for understanding the environmental
dependence of accretion activity as radio observations
overcome obscuration.  The Faint Images of the Radio Sky at Twenty
centimeters survey \citep{bec95} offers the advantage of providing
high angular resolution and sensitivity, and therefore is well suited
for extracting information regarding galactic nuclear activity.

We cross-matched the SDSS void and wall galaxy samples described in
Section ~\ref{data} with the FIRST catalog of sources with flux
densities exceeding 1 mJy at 1.4 GHz.  The global census of SDSS void
and wall galaxies that show radio activity at this level is
surprisingly different in voids and walls.  Table ~\ref{radiodata}
lists the results of cross-matching statistics and a comparison of the
$\nu L_{\nu}$(1.4 GHz) measurements per galaxy type.  Only 16 (1.6\%)
of void galaxies are brighter than 1 mJy at 1.4 GHz and they are all
strong line emitters.  Within walls, the fraction of objects with flux
densities exceeding this limit is significantly higher: 308 objects
(or 2.5\%), about 80\% of them showing line-emission activity.  These
calculations indicate at greater than 2-$\sigma$ that radio activity
in the centers of galaxies is less common in voids than in denser
environments.

This trend is not equally shared by galaxies of different spectral
properties.  We note that none of the void Ls are detected in FIRST
while a fraction as high as 7\% of the wall Ls appear as FIRST
sources.  Seyferts and Ts show very similar, if not identical
detection rates in FIRST in voids and walls; their corresponding
fractions of FIRST detections account for 15\%--20\%, and 9\%,
respectively.  Radio active H{\sc~ii}s are slightly more common in
walls than in voids.  While the difference in the FIRST detection rate
between all void and wall galaxies is statistically significant,
the variation of radio activity per spectral type remains
ambiguous, particularly for the void AGN where the number statistics
remain low.  If a single void LINER were detected in FIRST, the
fractions of void and wall Ls with radio activity would become
indistinguishable within the errors (5\% within voids versus 7\%
within walls).  
The level of radio activity also shows differences between the void
and wall systems, with the wall ones being clearly more luminous at
1.4 GHz.  A per type comparison remains however pointless because of
the very small number of void galaxies with measurable nuclear radio
emission.  We present, for reference, the $\nu L_{\nu}$(1.4 GHz)
luminosities in Table ~\ref{radiodata} as well.  Note also that none
of the void galaxies considered here would be considered radio loud;
the highest radio luminosity among void galaxies is $\nu L_{\nu}$(1.4
GHz) $= 1.7 \times 10^{38}$ erg s$^{-1}$, or $L_{\nu}$(1.4 GHz) $= 3
\times 10^{22}$ W Hz$^{-1}$.

These measurements are consistent with previous investigations of the
large scale structure of radio sources, but reveal potential new
features as well.  In particular, the fact that radio activity is
enhanced in wall regions relative to voids is in line with evidence
that radio loud galaxies are locally restricted to the super-galactic
plane, avoiding the voids (e.g., Shaver \& Pierre 1989).  However,
that the radio emitting H{\sc ii} systems are less common, and
probably weaker in voids than in denser regions, is at odds with
previous indications of a diminution in the star forming activity with
increasing local density, for both radio and optically selected
star-forming galaxies\citep{lew02, gom03, bal04, chr05, bes04, doy06}.
Note that while these studies examine the overall star-formation
activity within galaxies, either through galaxy colors or H$\alpha$
equivalent widths, our comparison refers to this kind of activity in
centers of galaxies only.  Thus, taken at face value, the discrepancy
shown by our measurements might indicate that core star-formation
activity shows somewhat opposite environmental dependence trends
relative to that occurring in the envelope; in lower density regions,
star-formation activity seems weaker in the core but stronger in the
envelope of galaxies.  That color gradients might also depend on
galaxy density, probably as strongly as star formation rate does, in
the sense that galaxies in rarefied regions exhibit more bluer
envelopes than those in crowded environs \citep{park07}, appear to
support this idea.

A possible explanation for this difference may be that the radio
emission associated with nuclear star-formation activity is not
entirely due to stellar activity, but it includes an additional
component.  A possible such contribution may be an AGN that is heavily
obscured, or its emission-line characteristics are swamped into the
light of the host galaxy or the circumnuclear star-forming regions;
note that the definition of the H{\sc ii} class relies on {\it
optical} line emission properties, and thus is not sensitive to
obscured accretion.  That AGN show possibly stronger radio emission in
walls than in voids might thus suggest that the enhanced radio
detection and emission in the wall H{\sc~ii}s comes from an accretion
source that is heavily obscured.  Although not detected in
optical wavelengths, such an AGN would be revealed at radio
frequencies.  This picture seems supported by other studies; e.g.,
hydrodynamical simulations of galaxy evolution processes, through
merger events, predict that for a large fraction of the accretion
time, the AGN (quasar) is heavily obscured, and that the intrinsic
quasar luminosity peaks during an early merging phase, but it is
completely obscured in optical \citep{hop06}, and galaxy spectral
analyses indicate that much of the BH growth occurs in AGN with high
amounts of dust extinction \citep{wild07}.  We explore this idea
further in the next sections where we compare accretion rates,
obscuration level, and stellar properties of void and wall systems of
different optical classifications.

\section{\sc Nuclear Activity and Host Properties} \label{voidaccretion}

We explore here common traits and differences in the relation between
nuclear emission-line activity and the host star-formation that void
and wall galaxies of various spectral types may reveal.  We gauge the
properties of the void and wall emission-line galaxies by means of
nuclear accretion rate, amount of fuel, and properties of their host
stellar population.  The parameters used in quantifying these
characteristics employ emission-line measurements and calculations
described by \citet{kau03b, kau03c}.  Our analysis is in some ways
similar to \citet{kau04}, however, we concentrate here specifically on
the least populated regions in the universe, the voids.  This work
constitutes the first in-depth scrutiny of Seyferts, LINERs,
Transition objects, and H{\sc~ii}s in these extreme environments.

\subsection{Accretion and Availability of Fuel} \label{accretion}

Following \citet{con06a}, we investigate the accretion power in the
wall and void active galaxies in terms of both $L_{\rm [O I]}$ and
$L_{\rm [O III]}$.  While the [\ion{O}{3}]$\lambda 5007$ emission is
susceptible to stellar contamination, the [\ion{O}{1}] $\lambda 6300$
line appears negligibly affected by this; in systems where
star-formation dominates the ionization process, i.e. the H {\sc
ii}s, [\ion{O}{3}] is one of the most prominent features while
[\ion{O}{1}] is hardly measurable.  The [\ion{O}{1}] and [\ion{O}{3}]
line fluxes are not simply a scale down (or up) of one another, the
[\ion{O}{1}]/[\ion{O}{3}] line flux ratio is significantly higher in
AGN than in other line-emitting systems \citep{hec80}.  Thus,
particularly for H{\sc~ii} and Transition objects, a comparison of
the $L_{\rm [O I]}$ and $L_{\rm [O III]}$ based estimates of central
accretion activity is an effective way of assessing the validity of
these measures.  Section \ref{o1o3} discusses this issue in detail.
As proxy for the accretion rate we measure the Eddington ratio as
quantified by log ($L_{\rm [O I]}$/${\sigma}^4$) and log ($L_{\rm [O
III]}$/${\sigma}^4$), similar to \citet{kew06}.  We gauge the
availability of fuel through estimates of the amount of intrinsic
obscuration, expressed in terms of Balmer decrements
H$\alpha$/H$\beta$, and the density of the emitting gas $n_e$, as
given by the [\ion{S}{2}]$\lambda$6716/$\lambda$6731 line ratio
\citep{ost89}.  We compare these parameters for void and wall galaxies
of different spectral types.

\begin{figure}
\plottwo{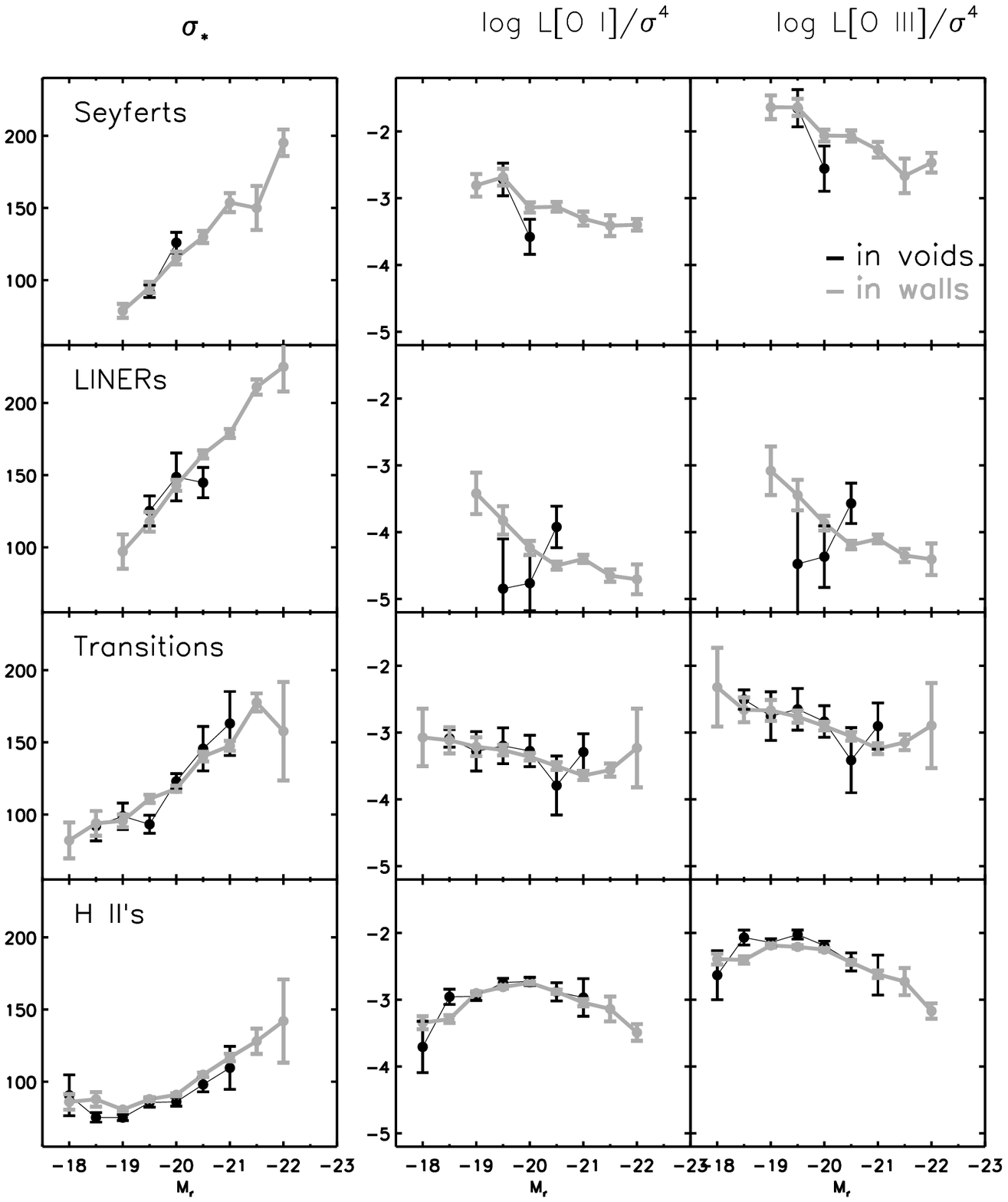}{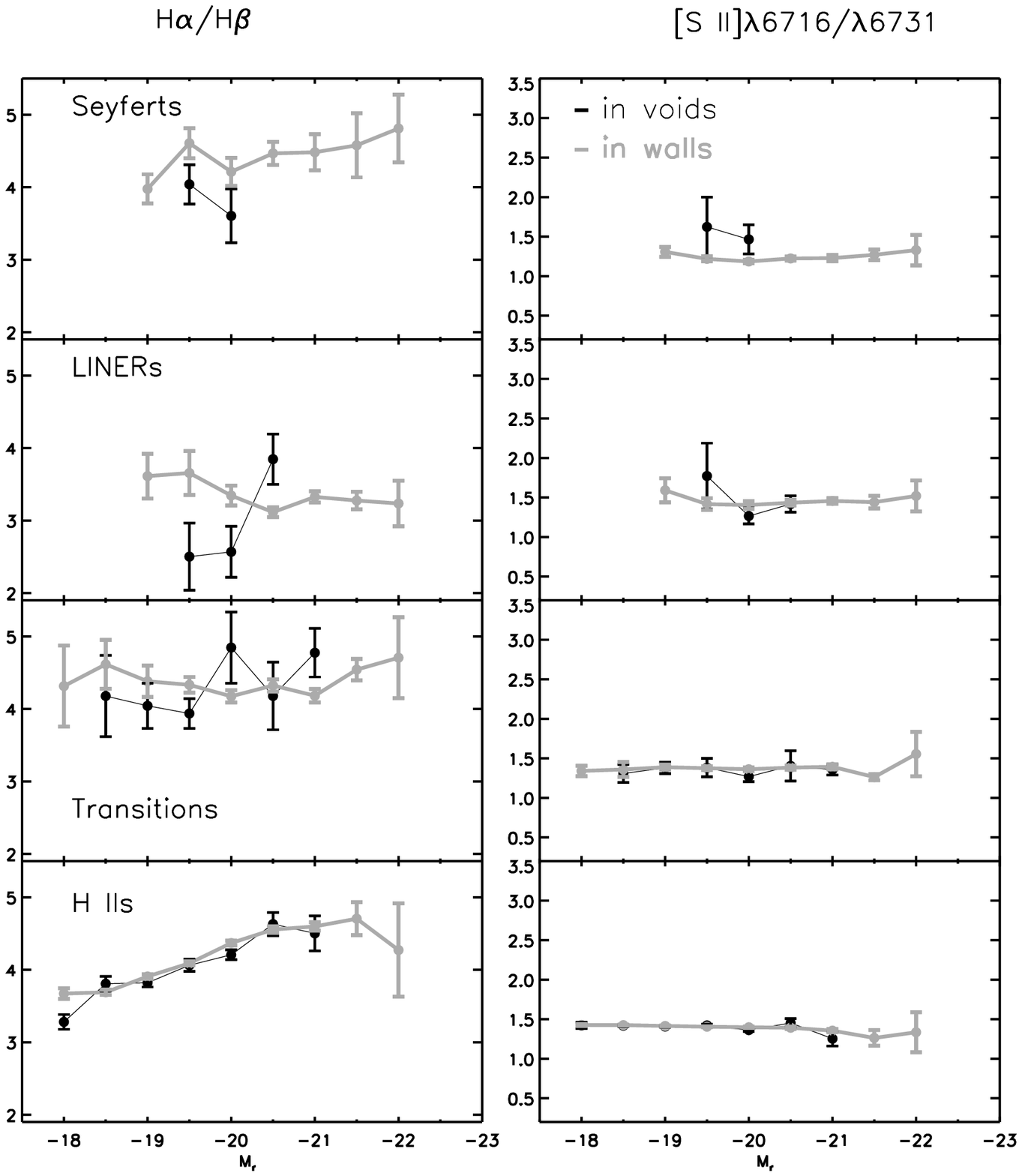}
\caption{ Mean values of $\sigma_*$, log $L_{\rm [O I]}/\sigma^4$ and
  log $L_{\rm [O III]}/\sigma^4$ (two proxies for the accretion rate),
  the Balmer decrement H$\alpha$/H$\beta$, and the
  [\ion{S}{2}]$\lambda$6716/$\lambda$6731 line ratio, measured in bins
  of 0.5 mag of $M_r$, for galaxies in voids (black) and walls (gray).
  The error bars represent the standard deviation of the mean.  Data
  are shown only for bins that include more than 2 objects.  Each row
  of plots represents a different subclass of objects, as indicated.
  The line luminosities have been corrected for internal dust
  attenuation, and are expressed in erg s$^{-1}$; the stellar velocity
  dispersions $\sigma_*$ are in km s$^{-1}$.
\label{o1o3nhne}}
\end{figure}

Figure ~\ref{o1o3nhne} illustrates how mean values in all of these
parameters, calculated in bins of $M_r$, compare in void and wall
galaxies of various spectral classifications.  It is interesting to
note the different environmental behavior that AGN hosted by bright
($M_r < -20$) and moderately bright ($M_r \ga -20$) galaxies exhibit.
Among the moderately bright hosts, both LINERs and Seyferts tend to
have lower accretion rates in voids than in walls, while in bright
galaxies, on the other hand, the small fraction of void AGN activity
(almost exclusively LINER-like) exhibits higher accretion rates.
Another interesting feature revealed by this plot is a potentially
close relation between the rates of accretion (i.e., Eddington ratios)
and the surrounding gas and obscuration: lower accretion rates in void
AGN are invariably associated with lower Balmer decrements and gas
densities, while higher accretion rates appear whenever the Balmer
ratios are higher.  While the statistical significance of each of
these trends is only of order 1-$\sigma$, there is a quite obvious
phenomenological continuity: these statistically-independent
measurements systematically follow the pattern expected if the
accretion rate correlates with the amount of dust and gas in their
hosts, and thus probably with that of material available for
accretion.


The boost in accretion in the LINERs inhabiting brighter galaxies is
an interesting finding and it is worth investigating its origin.  
Greater availability of accreting material can surely trigger such
behavior.  However, the origin of such an enhancement is not obvious,
given that their fainter counterparts exhibit the opposite effect.
One possibility would be that these systems have more material
available for accretion simply because matter is more abundant here.
Objects with stronger gravitational potential fields should be more
capable of keeping the matter inside, against galactic winds built by
the thermal pressure from supernovae, while low-mass systems would
eject their gas.  Thus, a look at the mass of their inner black holes
($M_{\rm BH}$) should shed some light on this issue.  Interestingly,
the comparison of their stellar velocity dispersions ($\sigma_*$), as
a measure of their $M_{\rm BH}$ (since $M_{\rm BH} \propto \sigma_*$,
e.g., Ferrarese \& Merritt 2000; Gebhardt et al. 2000) shown in Figure
~\ref{o1o3nhne} reveals that the void LINERs hosted by brighter
galaxies have systematically smaller mass BHs than similar wall
galaxies.  Thus, these particularly active accreting void systems do
not have, as conjectured, higher binding energies, and consequently,
they are not more powerful in cradling the material available for
accretion.  Another sensible way to produce the excess of material
(i.e., dust) in or around the LINERs hosted by bright void galaxies is
star formation.  We explore this idea in the next two sections, both
by comparing the [\ion{O}{1}] and [\ion{O}{3}] line emission and
through measures of the age of the stellar population in these
sources.

\subsection{[\ion{O}{1}] vs. [\ion{O}{3}] } \label{o1o3}


Because the [\ion{O}{1}] and [\ion{O}{3}] line emission mechanisms are
differently affected by star-forming and accretion activities, the
differences we see in these parameters between void and wall
line-emitting galaxies may suggest some interesting connection between
the two sources of ionization.  Although [\ion{O}{3}] is a very common
emission feature in star-forming galaxies, for the most luminous
systems (log $L_{\rm [O III]}$/erg/s $\ga$ 40.5) this parameter
appears to be a good measure of nuclear activity \citep{kau03, hec04}.
However, with the exception of very low $L_{\rm [O III]}$ objects,
which are LINERs \citep{con06a, kew06}, any other sample characterized
by log $L_{\rm [O III]}$/erg/s $<$ 40.5 is heavily contaminated by
objects with line flux ratios that better match an H{\sc~ii}-like
ionization than an accreting power source \citep{con06a}.  On the
other hand, [\ion{O}{1}] is extremely weak in systems whose ionization
is dominated by star-formation, thus [\ion{O}{1}] emission is an
important indicator of excitation due to accretion.  The reason is
that the [\ion{O}{1}] emission line arises preferentially in a zone of
partly ionized hydrogen, which can be quite extended in objects
photoionized by a spectrum containing a large fraction of high-energy
photons, i.e., originating in accretion, but is nearly absent in
galaxies photoionized by young, hot (OB) stars.  Therefore, more
vigorous star-formation activity would cause stronger [\ion{O}{3}]
emission, but should barely affect [\ion{O}{1}].  Correspondingly, the
[\ion{O}{1}] emission is expected to be enhanced in systems where
accretion is stronger.

Interestingly, the analysis of the [\ion{O}{1}] and [\ion{O}{3}] line
luminosities that we present in Section ~\ref{accretion} shows that
these two parameters compare remarkably well, revealing exactly the
same trends: both $L_{\rm [O I]}$ and $L_{\rm [O III]}$ are slightly
lower in voids than in walls in approximately the same amounts ($0.3$
dex for Seyferts and $0.5$ to $0.9$ dex for fainter Ls).  The standard
deviation of the mean in log $L_{\rm [O III]}$ is, however, generally
larger than that in [\ion{O}{1}], thus diluting the significance of
the trends in log $L_{\rm [O III]}$ with $M_r$.  The larger scatter in
$L_{\rm [O III]}$ is probably caused by the contribution of
star-forming activity to this line.  Such a contribution is expected
to be even more enhanced in void galaxies as they exhibit higher
specific star formation rates than their wall cousins
\citep{rojasspectro}.  Note also that for the LINERs hosted by void
bright galaxies, which show also evidence for younger stellar
populations (see Section ~\ref{sp}), the enhancement in $L_{\rm [O
III]}$ is larger than that in $L_{\rm [O I]}$; such a difference
signals again contribution by star-forming activity to the
[\ion{O}{3}] line emission.  Thus, particularly for void galaxies,
$L_{\rm [O III]}$ must be used with caution when employed in
estimating the accretion rate.

\subsection{Relation to Star Formation} \label{sp}


A variety of mechanisms have been suggested by which starbursts and
AGN are physically related.  An interesting idea is that the collapse
of very massive stars can give rise to the seed black hole, which can
be fed by stars either directly through tidal capture or indirectly
through gas shed via stellar mass loss.  Over time, the BH and its
host galaxy grow in size, but probably not in lock-step.  Because
more massive galaxies are less likely to harbor young stars
\citep{kau04}, and have more massive BHs \citep{mag98, geb00, fer00},
it is believed that the mass of the BH might have direct consequences
for star birth.  Hydrodynamic simulations of galaxy mergers including
black hole growth and feedback \citep{dimat05, hop05, cro06}
demonstrate how energetic outflows caused by gas falling onto the
central BH might self-regulate both BH growth and star formation in a
galaxy by heating the available gas and blowing it out, as previously
predicted by \citet{silk98}, or heating it up to temperatures that
suppress clumping and collapsing into stars.  Further star-formation
is therefore repressed as long as $M_{\rm BH}$ remains above a
critical value \citep{sch06}.

\begin{figure}
\plotone{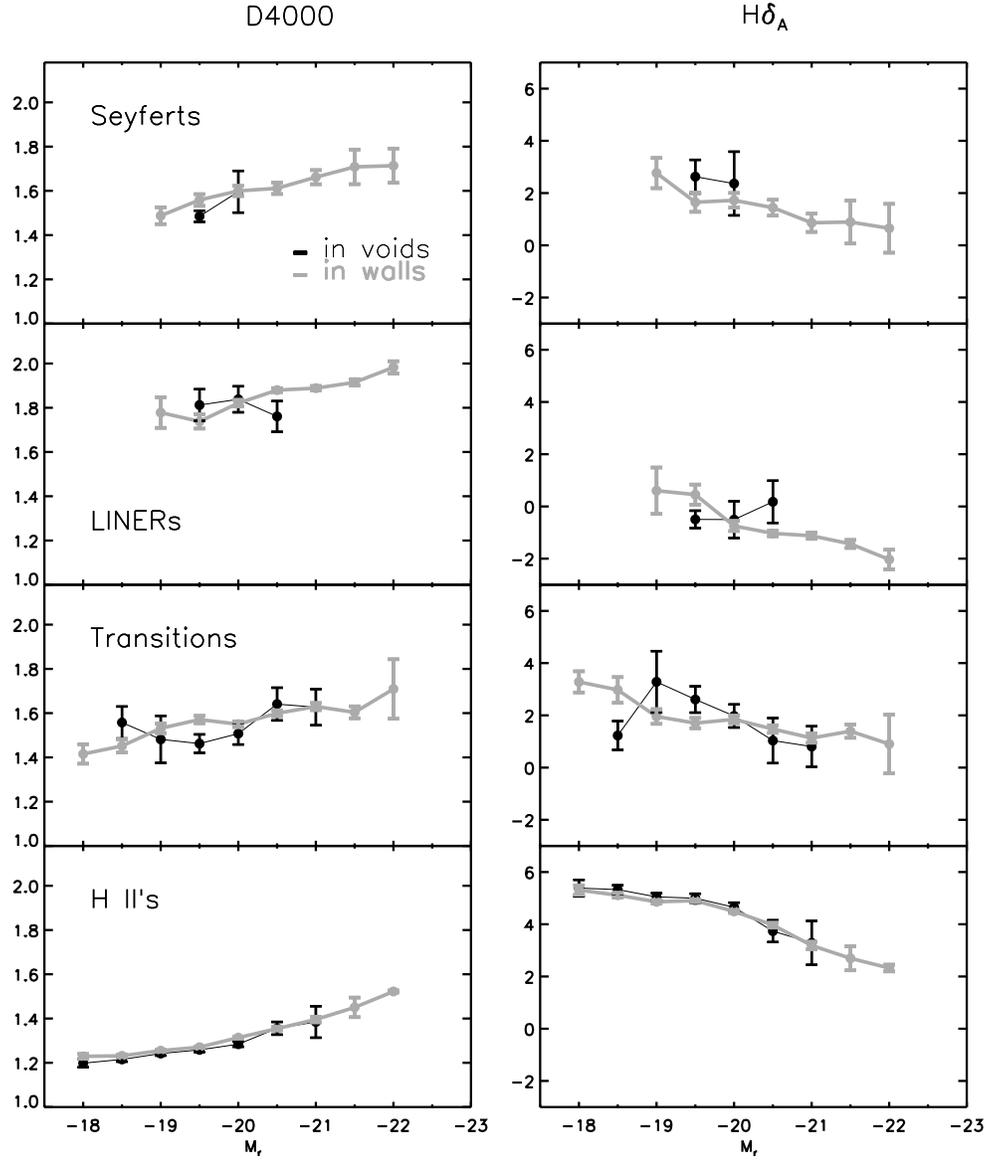}
\caption{ Mean values of the 4000\AA\ break ($D4000$) and the
  equivalent width of the H$\delta$ absorption line (H$\delta_A$), two
  independent measures of the mean stellar population ages.  Symbols
  and errors are as in Figure ~\ref{o1o3nhne}.  Note that with the
  exception of LINERs hosted by bright void galaxies which have
  younger stellar populations than their wall counterparts (smaller
  $D4000$ and larger H$\delta_A$ values), the star-formation histories
  in void and wall galaxies are very similar.
\label{stellarmag}}
\end{figure}

The regions least prone to galaxy-galaxy interactions or other
disturbing events, the voids, are some of the best test beds of this
scenario.  We proceed in this section with a comparative analysis of
the stellar population ages for various types of actively
line-emitting galaxies in voids and walls.  The strength of the
4000\AA\ break ($D4000$) and the equivalent width of the H$\delta$
absorption line (H$\delta_{\rm A}$) provide constraints on the mean
stellar age of a galaxy and the fraction of its stellar mass formed in
bursts over the past few Gyr \citep{kau03b}.  For the sake of
comparison with previous work \citep{kau03c,kau04,kew06}, we use the
measurements of the stellar characteristics derived and described by
\citet{kau03b} and \citet{bri04}.  Figure ~\ref{stellarmag} shows the
results of this comparison, in bins of $M_r$.

Globally, the void systems exhibit younger collections of stars than
those in walls: on average, $D4000$ is similar or slightly higher in
walls than in voids regardless of the types of objects compared,
however H$\delta_A$ is significantly lower (by $\sim 30\%$ in
Seyferts, and $\sim 60\%$ in Ls) in the more crowded environments than
in the underdense ones.  This comparison supports previous indications
of systematically younger galaxies in lower-density regions
\citep{kau04}.  Here we find that these trends extend to the most
extreme underdense environments as well.  We also show here that such
differences are due to the different range of host properties spanned
by the samples that are compared: the younger mean stellar ages of
void systems is a consequence of the general shortage of massive
bright galaxies (i.e.  the usual hosts of old stellar populations) in
the underdense regions.  The shift in the void luminosity function
toward lower brightness galaxies is nicely documented in
\citet{hoy05}.

At fixed host brightness, the properties of the central stellar
populations of void and wall galaxies are statistically similar,
however, there are exceptions.  The brightest ($M_r < -20$) void hosts
of LINERs stand out as interesting peculiarities, as they are clearly
associated with younger stellar populations (larger H$\delta_{\rm A}$,
smaller $D4000$) than those in walls.  Recall that these void LINERs
also show a peculiar boost in both accretion and amount of
obscuration, which might indicate greater availability of fuel.
Similarly, the fainter ($M_r \geq -20$) void hosts of LINERs, which
show weaker accretion activity and significantly less obscuration,
also show signs of older stellar emission (lower H$\delta_{\rm A}$
values).  All these trends indicate that the strength of AGN activity
is correlated with the age of the associated stellar population, in
the sense that AGN activity is stronger when the surrounding stars are
relatively young (few Gyr old, corresponding to $D4000 \la 1.6$, and
H$\delta_{\rm A} \ga 2$), while weak AGN activity is associated with
mean stellar ages that are invariably older than 10 Gyr ($D4000 \ga
1.8$, and H$\delta_{\rm A} \la 0$).  In other words, AGN activity and
star-formation are somewhat coeval, and it is possible that the same
reservoir of fuel is used in igniting these activities.  Moreover,
that the weak AGN harbor generally heavier (grown up) BHs and that
they are routinely associated with lack of obscuration, low gas
densities, and older stellar populations, may well be the consequence
of the fact that star formation remains suppressed once the BH mass is
large enough.  Such trends are somewhat apparent among galaxies in
general, however this comparison of the emission-line void and wall
galaxies reveals them more explicitly.

\section{\sc The Small Scale Environment: Nearest Neighbor Statistics} 
\label{nnprop}

A key point in investigating AGN in voids is understanding at what
scale, if at all, the environment affects accretion activity.  As
shown by \citet{con06a}, different types of galactic activity prefer
different large scale structures; LINERs are most clustered, H {\sc
ii}s are the least clustered, Seyferts and Ts are less clustered than
Ls but more clustered than H{\sc~ii}s.  Dark matter density
fluctuations are however important at all scales.  In fact, because
the fractional density fluctuations are largest on the smallest
scales, any effect that depends on density will appear to be most
strongly correlated with smaller than with larger scales.  One would
thus expect that the small scale climate also influences the strength
and type of accretion or star-formation, and even the link between
them.  

The rate of interaction is supposedly reduced in voids, and thus, an
analysis of the effects of small scale environs on galactic activity
is particularly instructive.  In particular, it is of interest to find
out whether objects displaying a certain type of activity are located
in the highest or lowest density substructures within the most
underdense environs, and whether AGN are associated or not with small
groups within voids.  To investigate these issues we measure the
distance to the third nearest neighbor ($d_{\rm 3nn}$) as a density
indicator, and the distance to the first nearest neighbor ($d_{\rm
1nn}$) as a measure of the probability of having close encounters.  We
measure $d_{\rm 1nn}$ separately for the volume limited sample, which
thus counts only the bright ($M_r < -19.5$) neighbors, and for the
whole (flux limited) sample, which adds fainter galaxies.  We note
here that a small number of close neighbors are missed due to the
55$\arcsec$ minimum fiber separation; this fiber collision issue is
however unlikely to affect significantly our comparative analysis of
near neighbor statistics of void and wall samples as the effects
should be nearly identical for all samples involved, and thus should
cancel out.

\begin{figure}
\plotone{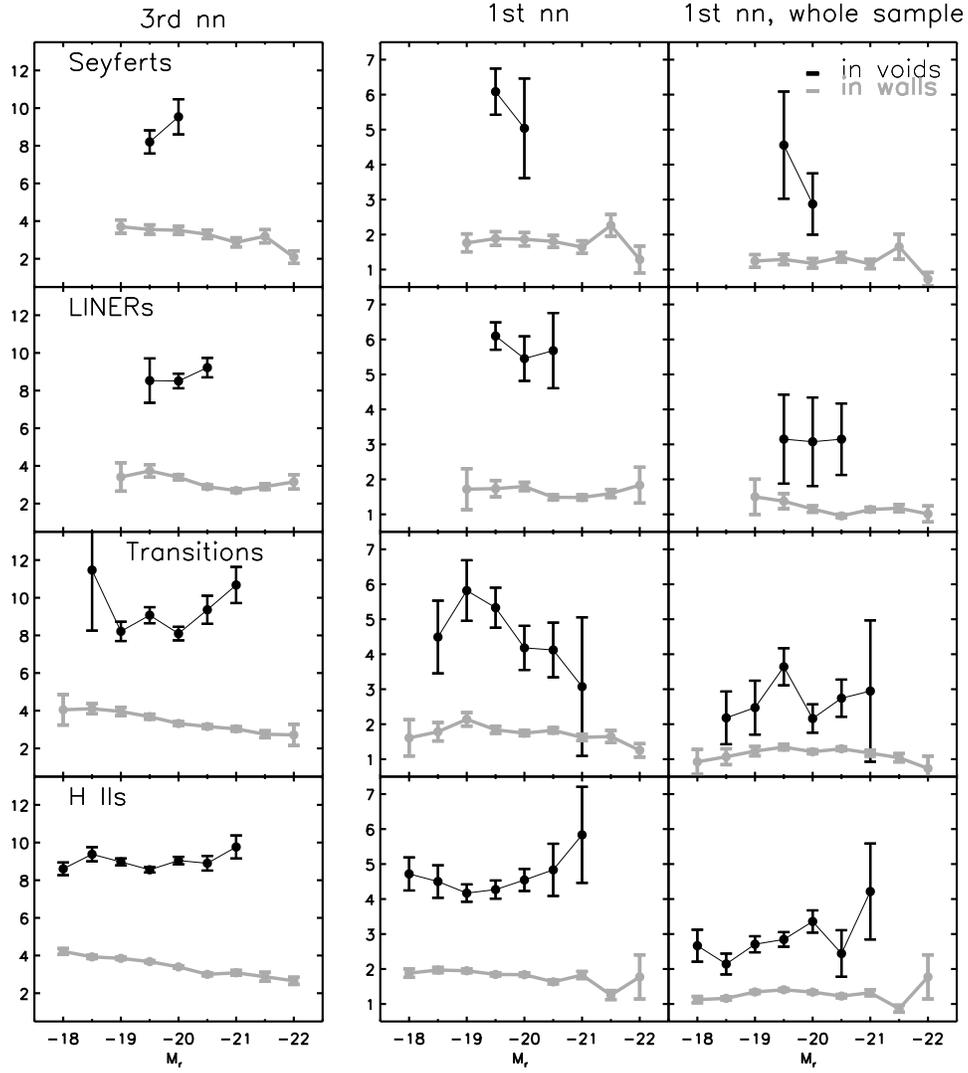}
\caption{ The 3rd and 1st nearest neighbor statistics averaged in
$M_r$ bins.  Symbols and colors are as in Figure ~\ref{o1o3nhne}.  The
tendencies shown by the median values are preserved for all brightness
ranges: LINERs are furthest from other bright galaxies in voids, but
have the nearest neighbors when they live in walls.
\label{nnmag}}
\end{figure}

Table ~\ref{nn} lists the median values in these parameters for both
void and wall systems of various spectral types, and Figure
~\ref{nnmag} illustrates how these parameters compare at fixed $M_r$.
Because the definition of voids and walls employs $d_{\rm 3nn}$, all
void galaxies have larger $d_{\rm 3nn}$ than those in walls.  As
expected, $d_{\rm 3nn}$ varies little among void galaxies\footnote{The
fact that we only see small variation in $d_{\rm 3nn}$ in voids is not
surprising.  In these regions, we are measuring distances $ > 7
h^{-1}$ Mpc, and hence, we are looking at quite large smoothing scales
where the $rms$ density fluctuations are smaller than on the scales
where $d_{\rm 3nn}$ is measured in walls.}, the median values are
consistent with each other within $1-\sigma$.  It is interesting to
note however an apparent tendency of increasing $d_{\rm 3nn}$ with
host luminosity (Figure ~\ref{nnmag}); this suggests that, in voids,
more luminous and therefore massive systems are less grouped than the
fainter, less massive ones.

Among the wall galaxies, $d_{\rm 3nn}$ shows significant differences
among objects of different spectral types, and trends that are {\it
opposite} to those exhibited by void systems: LINERs have the closest
3rd neighbor (with an average of $3.0 \pm 0.1$ $h^{-1}$ Mpc), and thus
the highest density, while the H{\sc~ii}s populate less dense
regions within walls (their average $d_{\rm 3nn} \sim 3.6$ $h^{-1}$
Mpc).  Seyferts and Ts show very similar values ($d_{\rm 3nn} \sim
3.3$ $h^{-1}$ Mpc), which are intermediate between LINERs and H {\sc
ii} galaxies.  These trends persist even when the samples are split in
bins of $M_r$, and it is pretty clear that $d_{\rm 3nn}$ decreases
with $M_r$ (Figure ~\ref{nnmag}).  Such findings are consistent with
the difference in clustering that \citet{con06a} reveal, with LINERs
being the most clustered active galaxies, Seyferts showing lower
clustering amplitude and H{\sc~ii}s being the systems that are the
least clustered.  Note however that the clustering analysis involves
auto-correlations of each type of these emission-line objects, while
the nearest-neighbor analysis cross-correlates each type of object
with the full population of galaxies.

Distance to the nearest galaxy, $d_{\rm 1nn}$, may be important in
assessing whether or not interactions are important in triggering
nuclear activity in galaxies.  The $d_{\rm 1nn}$ measurements
generally reinforce the behavior shown by $d_{\rm 3nn}$ but also show
some peculiar trends.
When measured within the volume limited sample, $d_{\rm 1nn}$
indicates that, in voids, LINERs and Seyferts are closer to their
bright neighbors than the H{\sc~ii}s while Ts show a somewhat
intermediate behavior.  Kolmogorov-Smirnov tests are consistent with
clear differences in the distributions of H{\sc~ii}s and LINERs, (the
probability that the $d_{\rm 1nn}$ distributions of these two samples
share the same parent population is Prob$_{\rm H II-L} = 0.017$).  When
fainter neighbors are counted, in the magnitude limited sample, the
likelihood of having close neighbors is very similar for all types of
void emission-line systems.  In walls, LINERs have the closest
neighbors, both faint and bright, while H{\sc~ii}s are the farthest
away from any galaxy.  These trends suggest that actively star-forming
systems prefer the densest sub-regions in voids and the most rarefied
neighborhoods within crowded environs, while the systems in which
accretion is strong enough to be detected is generally associated with
a grouped small-scale environment.

The comparison of these distances in bins of $M_r$, as shown in Figure
~\ref{nnmag}, may explain these particular trends.  For all three
measures of density, and in particular for $d_{\rm 3nn}$, the
differences between void and wall galaxies of given brightness is
greater for the more luminous objects: the brighter a galaxy is, the
more isolated it seems to be in voids, while the most ``connected'' in
walls.  Along these trends, H{\sc~ii}s, which are generally fainter
than other types of objects, occupy more crowded void sub-environments
but more rarefied regions in walls, while LINERs, which inhabit
generally bright hosts, are the most isolated in voids and the most
grouped in walls.

That more luminous galaxies appear to prefer more crowded regions
within walls is consistent with previous results that more luminous
systems are more clustered (on large scale).  However, their tendency
to be more isolated within voids is a new and surprising finding.
This observation suggests that while the small scale environment may
influence the intrinsic properties of their inhabitants (e.g., $M_r$),
the dynamics that determine the large scale structure (e.g., expansion
versus contraction) are equally important as they influence the
likelihood for interaction, and thus the type of galactic nuclear
activity.

It is important to note that both $d_{\rm 1nn}$ and $d_{\rm 3nn}$
address the embedding of the AGN hosts and other galaxies in dark
matter halos, however, at different scales: $d_{\rm 1nn}$, of order
1.5 Mpc h$^{-1}$ in walls and about three times higher in voids,
examines neighbors that may lie within the same dark matter halo,
while $d_{\rm 3nn}$, which is $\sim$3 Mpc h$^{-1}$ in walls and
$\sim$8.5 Mpc h$^{-1}$ in voids, measures the large scale environment
traced by multiple halos.  Uncertainty in the $d_{\rm 1nn}$ measure
arises because of (1) the fiber collision problem and (2) peculiar
velocities along the line of sight.  Both effects will tend to
increase $d_{\rm 1nn}$ by removing a few close neighbors and by
spreading virialized systems along the line of sight,
respectively. Thus, the $d_{\rm 1nn}$ values in walls, where $d_{\rm
1nn}$ is small enough to be affected by both systematics, may be
overestimates. In voids, these systematic effects go in the same
direction, however are likely to be smaller, because the density is
much lower.  The $d_{\rm 3nn}$ measurements are likely to be affected
by these biases in a similar manner, however much less.  These
arguments indicate thus that our conclusions can in fact be stronger.
In particular, the differences in $d_{\rm 1nn}$ could certainly not be
caused by systematic effects, and are generally underestimated.
Moreover, the comparison between types of active galaxies in voids and
walls should be hardly affected, as the spectral classification and
near neighbor statistics are completely independent analyses.

To summarize, the youngest, actively star-forming, fuel-rich systems,
i.e., the H{\sc~ii}s, appear in weakly grouped environments, that are
the densest regions within voids and the rarefied ones within walls.
On the other hand, the most common type of AGN activity (i.e.,
LINER-like), which is relatively feeble and associated with old
massive evolved hosts, shows a peculiar preference for the most
clustered wall substructures and the emptiest among voids.  This trend
argues against a major role played by close encounters or
mergers in triggering or sustaining their activity.  Interestingly,
the most active AGN (Seyferts and Ts) prefer the relatively low
density wall sub-environments and the somewhat grouped void ones; this
is a behavior that is intermediate between those manifested by LINERs
and H{\sc~ii}s.


\section{\sc Conclusions and Discussion} \label{discussion}

\subsection{\sc Summary of Results} \label{results}

We use the largest sample of voids and void galaxies yet defined to
investigate the galactic nuclear accretion phenomenon in the most
underdense regions, in relation to their more populous counterparts.
By employing spectroscopic and photometric data based on the SDSS DR2
catalog, and in particular measurements available in the Garching
catalog, we conduct a comparative analysis of void and wall systems of
different radiative signature. 

We find that all types of Low Luminosity AGN exist in void regions.
However, their occurrence rate and intrinsic properties show variation
from their wall counterparts.  The differences between the wall and
void AGN seem to be driven by the properties of their hosts, which are
correlated with (or governed by) their small scale environment.
Following is a summary of our main results:

(i) Among moderately bright or fainter galaxies ($10 < log
(M_*/M_{\sun}) < 10.5$, $M_r \ga -20$), the rate of occurrence of AGN
is higher in voids than in walls.  The most common accretion activity
in voids is of medium power, with $L_{\rm [O III]} \sim 10^{38}$ erg
s$^{-1}$.  For the more luminous massive hosts, due to small number
statistics, the relative prevalence of accretion activity in voids
versus walls remains poorly constrained.

(ii) The majority of void AGN, which are hosted by $M_r \ga -20$
galaxies, have the tendency to accrete at lower rates than in those in
walls. This behavior seems related to the fact that the void systems
show less obscuration and, perhaps, less dense emitting gas.  That the
stellar populations associated with void and wall AGN are similarly
aged suggests that fuel might be equally available for accretion in
void and wall galaxies of similar properties (i.e., $M_r$), but that
fuel is less efficiently driven towards the nucleus in void galaxies.

(iii) The few void AGN hosted by bright, massive galaxies ($M_r \la
-20$, log $M_*/M_{\sun} > 10.5$) are LINERs that show peculiarly
higher accretion rates, larger amounts of obscuring matter and more
recent star-formation than in their wall counterparts.  
These particular systems reinforce the general trends other objects
show:  higher accretion rates are invariably associated with younger
stellar populations and higher obscuration.    These trends suggest that
the amount of obscuration could be a measure of the available fuel
for both star formation and accretion.

(iv) The radio activity of line-emitting galaxies appears both less
frequent and weaker in voids than in walls.  Were we able to support
these differences with statistically significant measurements, they
would imply that central radio activity in wall systems, including H
{\sc ii}s, is more pronounced because it builds on contributions from
accretion that remains optically obscured and therefore undetected.

(v) Nearest neighbor statistics show that the type of emission-line
activity is correlated with the small-scale local environment.  The
star-forming regions (the H{\sc~ii}s), populate the most crowded
sub-regions of voids while populating relatively sparse regions in
walls; both are environments where low-mass galaxies recently formed.
The weakly active galaxies (LINERs) live within the clusters in walls
but the most rarefied regions in voids.  This finding is puzzling and
suggests that these systems, which are generally old, were probably
not aware of their environments when they formed.  Actively accreting
systems (Seyferts and possibly the Ts) inhabit intermediate regions,
which are relatively dense galaxy neighborhoods in voids but are of
average density in walls.

(vi) These correlations among the type and strength of galactic
nuclear activity, incidence rates of different types, and their small
and large scale environments, suggest an
H{\sc~ii}$\rightarrow$S/T$\rightarrow$LINER evolutionary scenario in
which interaction is responsible for propelling gas towards the galaxy
centers, triggering star formation and feeding the active galactic
nucleus.

\subsection{\sc The H{\sc~ii}$\rightarrow$S/T$\rightarrow$LINER
  Evolutionary Sequence} 
\label{sequence} 

\begin{figure}
\plotone{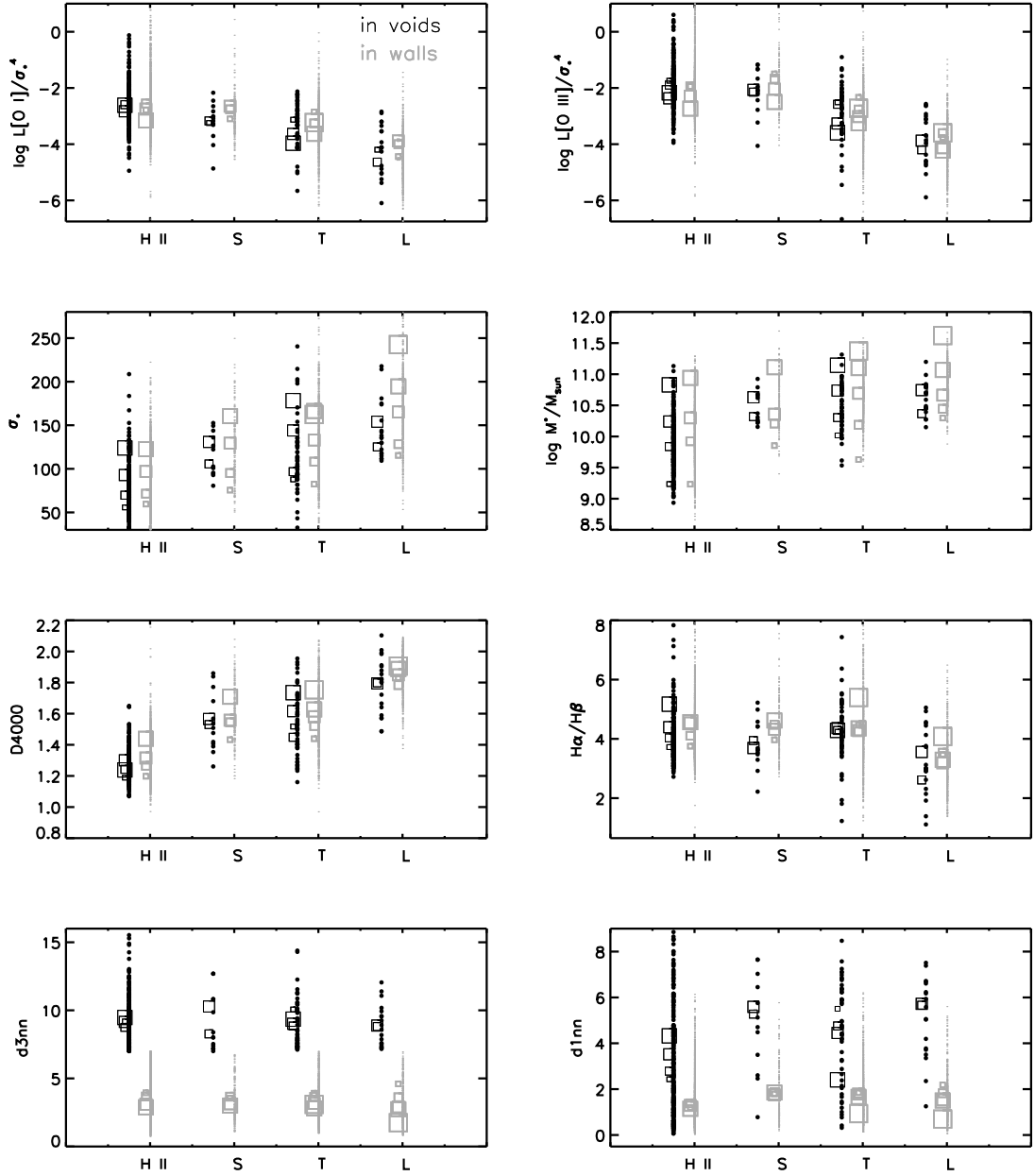}
\caption{ Trends of galaxy properties in voids (black) and walls
  (gray) along the H{\sc~ii}$\rightarrow$S/T$\rightarrow$L sequence described in the
  text: accretion rate, as expressed by $L/L_{\rm Edd}$, estimated
  based on both log $L[{\rm O I}]/\sigma_*^4$ and log $L[{\rm O
  III}]/\sigma_*^4$, BH mass, as given by $\sigma_*$, stellar mass,
  star formation rate (or youngness of the associated stellar
  population), and level of obscuration.  Data points represent
  individual measurements while square boxes indicate median values of
  subsets of objects separated in four bins in $M_r$; the size of each
  square scales with the respective median $M_r$.  The bottom panels
  show, for comparison, the distances to the nearest 3rd and 1st
  (bright) neighbors.
\label{h2stl}}
\end{figure}

Figure ~\ref{h2stl} illustrates how various intrinsic and host
properties of actively line-emitting galaxies follow this H {\sc
  ii}$\rightarrow$S/T$\rightarrow$LINER sequence.  
The early stages of such objects manifest 
themselves as H{\sc~ii}, as the accreting source remains heavily
embedded in dust.  As the star-burst fades in time, the dominance of
the Seyfert-like excitation in systems of generally small but actively
accreting black holes becomes more evident.  Successive evolution
reveals aging stellar populations associated with objects spectrally
classified as Transition objects, that are still showing signs of
accretion, followed by LINERs, whose stars are predominantly old and
whose accretion onto already grown-up BHs is close to minimal.  Note
that this H{\sc~ii}$\rightarrow$S/T$\rightarrow$LINER progression is
very similar in walls and voids.  
The lower accretion rates and the higher frequency
of actively accreting systems in void versus wall galaxies of similar
properties indicate a potential delay in the AGN dominance phase
within voids.  Thus, void AGN progress through the H
{\sc~ii}$\rightarrow$S/T$\rightarrow$LINER sequence more slowly, while
the sequence is similar for both 
void and wall galaxies.  This picture fits well the observed
properties of each type of galaxy nucleus:

\begin{itemize}
\item That H{\sc~ii}-type of activity is significantly more frequent
  in voids than in walls suggests that their void-like environments,
  in which they have closer (both 1st and 3rd) neighbors than other
  types of objects have, are essential in triggering their activity.
  In other words, close encounters that produce either harassment, or
  major and/or minor mergers may be an important cause for igniting
  both accretion and star formation.

\item Seyferts' environments in both voids and walls are intermediate
  between those of H{\sc~ii}s and LINERs, regardless of their
  brightness.  If the Seyfert-like activity is triggered by
  interactions, probably the same ones that turn on the H{\sc~ii}s,
  there must be a time lag between the onset of the star-burst and
  when accretion becomes dominant, or simply observable.  Such a time
  interval corresponds to a period of aging of the stellar population
  (as seen in the differences in $D4000$ and H$\delta_{\rm A}$ between
  H{\sc~ii}s and Seyferts), when the post starburst fuel becomes
  increasingly available for accretion.  Moreover, this progression
  develops relatively uninterrupted in voids, and therefore, possibly,
  at a slower pace than it would in walls where close encounters or
  other types of interactions can either accelerate or terminate it.

\item Void and wall Ts are barely distinguishable in their physical
  characteristics.  In both voids and walls, their nn-statistics and
  intrinsic and host properties are intermediate between those of
  LINERs and H{\sc~ii}s, for any given range of $M_r$.  Their BHs are
  apparently growing (their accretion activity is stronger than that
  of Ls but weaker than that of Seyferts), their fuel supply seems
  plentiful (they are found in some of the most obscured systems), and
  they are associated with (quite massive) stellar populations that
  are generally younger than those of LINERs, but older than the
  majority of star-forming systems.  It is among Ts however that the
  most massive void accreting BHs are observed; this might suggest
  that, in the proposed H {\sc~ii}$\rightarrow$S/T$\rightarrow$LINER
  sequence, massive 
  void galaxies reach the low accretion rate (i.e., LINER) phase later
  than in walls.

\item Whether Seyferts or Transition objects are first in this
  sequence remains ambiguous.  Both their intrinsic properties and
  nn-statistics are very similar, and remain intermediate between
  those of H {\sc~ii}s and LINERs.  The H$\alpha$/H$\beta$ Balmer
  decrements and the $d_{\rm 1nn}$ are the only parameters that show a
  ``jump'' in the otherwise smooth
  H{\sc~ii}$\rightarrow$S$\rightarrow$T$\rightarrow$LINER  
  sequence manifested by other properties of these systems.  Further
  investigations of these objects should address the differences
  between them in terms of a possible evolutionary progression.

\item Although we do not provide any quantitative estimates of the
time spent in or during the various phases we propose here, this this
analysis shows that both void and wall galaxies follow the same cycle.
The AGN evolution does not affect the gravitational environment.  To
the contrary, it is the environment that sets the time scale for
evolution along such a sequence; it seems to take longer to march
through the different phases in voids than in walls, but the physics
is the same.  The large scale clustering is consistent with this
picture: LINERs are {\it now} more clustered because objects in dense
regions underwent the H{\sc~ii}$\rightarrow$S/T$\rightarrow$LINER evolution more
quickly; the higher rate of galaxy-galaxy interactions speeds up the
way AGN proceed through the sequence.  Hosts whose central regions are
now in the H {\sc~ii} phase will always be less clustered than current
LINERs.

\end{itemize}

Although far from being complete, this proposed evolutionary sequence
is engaging and offers a comprehensive picture for the co-evolution of
AGN and their host galaxies.  The broad idea that mergers trigger star
formation and that the AGN appears afterwards, in fact shutting off
the star formation because of feedback, has been discussed previously
in the literature.  For example, N-body simulations by \citet{byr86}
and \citet{her95} showed that interactions drive gas toward the
nucleus and can produce intense star formation followed by an AGN.
More recent state-of-the-art hydrodynamical models \citep{dimat05,
spr05, hop05, hop06} show that during mergers, the BH accretion peaks
considerably {\it after} the merger started, and {\it after} the
star-formation rate has peaked.  However, whether early bright quasars
and later, dimmer AGN obey similar physics needs still to be
addressed.  The H{\sc~ii}$\rightarrow$S/T$\rightarrow$L sequence that
this study reveals, based on the smooth alignment of several of their
spectral properties, may be the first empirical evidence for an
analogous duty cycle in high redshift bright systems and in nearby
galaxies hosting weak quasar-like activity.

This scenario can also accommodate the rather inconclusive findings
regarding the role of mergers in activating AGN: their hosts do not
show evidence for bars \citep{mul97, lai02} or disturbances caused by
galaxy-galaxy interactions \citep{mal98}, exhibit morphologies very
similar to those of field galaxies \citep{derob98a, derob98b}, and
pair counting in both optical and IR remain inconclusive as possible
excess of companions are sometimes found \citep{dah84, raf95}, but not
always \citep{fue88, lau95}.  Moreover, \citet{sch01, sch04} show that
claims of evidence of interactions in the literature could be
attributed to selection effects.  The
H{\sc~ii}$\rightarrow$S/T$\rightarrow$LINER cycle suggests that the
majority of AGN might be detected only a certain period after the
interaction, allowing time for the starburst to fade and for the BH
accretion to gain strength.

One might argue that that other forms of evolution could also exist
for these emission-line galaxies, as opposed to this simple
progression.  We note that this proposed sequence does not imply that
every HII galaxy at the present epoch must necessarily become a LINER;
it is certainly possible that some systems go through H{\sc~ii}-only
phases, or L-only phases, which might not be part of the larger
progression.  

We would however like to emphasize that the timescales
necessary to transform from one galaxy type to the next are quite
reasonable.  At a first glance, this is somewhat surprising given the
relatively large range in BH masses ($2 \times 10^7 M_{\sun}$ for
H{\sc~ii}s, to $2 \times 10^8 M_{\sun}$ for LINERs, inferred from
$\sigma_*$'s assuming, e.g., Tremaine et al. 2002), and their
significantly low accretion rates ($L/L_{\rm Edd} \la 0.005$; see
Figure ~\ref{h2stl}, and note that log $L[{\rm O III}]/\sigma_*^4 =
-1$ corresponds to approximately $L/L_{\rm Edd} = 0.05$, according to
Kewley et al. 2006); apparently, for a canonical value for the
accretion efficiency, i.e., 10\%, it takes approximately a Hubble time
to e-fold in BH mass.  The key is in the fact that the low luminosity
AGN accrete inefficiently, as very little energy generated by
accretion is radiated away (the optically thin cooling time of the gas
is longer than the inflow time).  Studies show that, in these cases,
the radiative efficiency can be as low as $10^{-6}$ \citep{ree82,
nar94, qua03}.  With a (not a necessarily extreme) efficiency value of
$10^{-5}$ then, the e-folding time in the BH mass is $\approx 4.5
\times 10^2/l$ years, and thus, as little as few Myrs for, e.g., $l =
L/L_{\rm Edd} = 0.0005$.  This (not necessarily extreme) example shows
then that such an evolutionary scenario is actually physically
possible.

The peculiarity of LINER environments is a new and intriguing result
and clearly shows that this type of activity is not controlled by its
surrounding environment.  Galaxies hosting LINERs could be associated
with high initial density peaks in the dark matter distribution, which
evolved subject to their large scale environment.  Being generally
massive systems to start with, LINERs' hosts would be prone to
accreting material around them.  In voids, this ``cleaning''
enterprise would contribute to emptying the already rarefied
neighboring space, leaving little or insufficient material for future
formation of (massive, bright) galaxies; they would thus end up in the
most underdense void neighborhoods.  In walls, and in particular
within clusters, accretion of surrounding material would make a small
difference as the matter density is higher.  Simulations of dark
matter halos and correlations with the properties of their inhabiting
galaxies should be able to address these ideas.


Further tests of this H{\sc~ii}$\rightarrow$S/T$\rightarrow$LINER evolutionary
scenario are clearly needed.  These tests require larger samples that
allow separation into different morphological subsamples, and
observables that parametrize the galaxy morphology better than the
concentration index.  Moreover, we need better constraints on the BH
masses and consequently the Eddington rates for the H{\sc~ii}s in
particular, or the late type galaxies in general.  When available,
analysis of such parameters would shed light on the assumed coevality
of star-formation and Seyfert-like BH accretion in centers of
galaxies, removing ambiguities regarding the initial stages of
the H{\sc~ii}$\rightarrow$S/T$\rightarrow$LINER progression.

\vspace{1cm}

SPECIAL NOTE: During the review process of this paper, we became aware
of another piece of work which introduces the same idea of an
evolutionary sequence, ``from star formation via nuclear activity to
quiescence`` \citep{sch07}.  Interestingly, although their general
approach, analysis, and samples used, are quite different from ours,
the measurements on which the evidence for a ``star-forming,
transition object, Seyfert, LINER'' sequence is built shares a common
set of parameters with those used in our analysis: stellar velocity
dispersions, ages of starbursts, and reddening.  The global frame in
which this evolutionary sequence in presented is however definitely
different.  While Schawinski et al. take a step forward toward
understanding this possible time sequence among early-type galaxies
and provide an in-depth analysis of the possible time scales involved
in this picture, our paper offers a broader perspective on how such an
evolutionary sequence constrains the galaxy evolution models as we
provide important links to the environment.


\vspace{1cm}

\acknowledgments Support for this work was provided by NASA through
grant NAG5-12243 and by NSF through grant AST-0507647.  The authors
thank John Parejko for making available his SDSS-FIRST galaxy
cross-matching catalog, and the anonymous referee for a helpful
report.
    
Funding for the SDSS and SDSS-II has been provided by the Alfred
    P. Sloan Foundation, the Participating Institutions, the National
    Science Foundation, the U.S. Department of Energy, the National
    Aeronautics and Space Administration, the Japanese Monbukagakusho,
    the Max Planck Society, and the Higher Education Funding Council
    for England. The SDSS Web Site is http://www.sdss.org/.

    The SDSS is managed by the Astrophysical Research Consortium for
    the Participating Institutions. The Participating Institutions are
    the American Museum of Natural History, Astrophysical Institute
    Potsdam, University of Basel, Cambridge University, Case Western
    Reserve University, University of Chicago, Drexel University,
    Fermilab, the Institute for Advanced Study, the Japan
    Participation Group, Johns Hopkins University, the Joint Institute
    for Nuclear Astrophysics, the Kavli Institute for Particle
    Astrophysics and Cosmology, the Korean Scientist Group, the
    Chinese Academy of Sciences (LAMOST), Los Alamos National
    Laboratory, the Max-Planck-Institute for Astronomy (MPA), the
    Max-Planck-Institute for Astrophysics (MPIA), New Mexico State
    University, Ohio State University, University of Pittsburgh,
    University of Portsmouth, Princeton University, the United States
    Naval Observatory, and the University of Washington.


\begin{deluxetable}{lcccccc}
\tablecolumns{7} \tablewidth{0pt}
\tablecaption{Object Sample Statistics.
\label{statistics}}
\tablehead{
\colhead{} &
\colhead{} &
\multicolumn{2}{c}{in voids} &
\colhead{} &
\multicolumn{2}{c}{in walls}\\
\cline{3-4} \cline{6-7}\\
\colhead{Sample} &
\colhead{} &
\colhead{$N_{obj}$} &
\colhead{Fraction\tablenotemark{a}} &
\colhead{} &
\colhead{$N_{obj}$} &
\colhead{Fraction\tablenotemark{b}}}
\startdata
Emission-line & & 424 &42.6\% & & 4183 &33.4\% \\
...Seyfert    && 13 & 1.5\% && 190 & 1.5\% \\
...LINER      && 20 & 2.0\% && 510 & 4.1\% \\
...Transition && 53 & 5.3\% && 796 & 6.4\% \\
...H II       &&327 &32.8\% &&2610 &20.8\% \\
\enddata
\tablenotetext{a}{Fraction of all 996 void galaxies.}  
\tablenotetext{a}{Fraction of all 12153 wall galaxies.}  
\end{deluxetable}


\begin{deluxetable}{lccccc}
\tablecolumns{7} \tablewidth{0pt} \tablecaption{ Host Properties
\label{host}}
\tablehead{
\colhead{} &
\multicolumn{2}{c}{$M_r$} &
\colhead{} &
\multicolumn{2}{c}{$C$\tablenotemark{a}}\\
\cline{2-3} \cline{5-6}\\
\colhead{Sample} &
\colhead{void} &
\colhead{wall} &
\colhead{} &
\colhead{void} &
\colhead{wall}}
\startdata
Seyfert    &~-19.8$\pm$0.1~&~-20.2$\pm$0.1~& &~2.8$\pm$0.1~ &2.67$\pm$0.03\\
LINER      &~-20.2$\pm$0.2~&~-20.7$\pm$0.04& &2.96$\pm$0.05 &3.00$\pm$0.02\\
Transition &~-19.7$\pm$0.1~&~-20.2$\pm$0.03& &2.67$\pm$0.04 &2.70$\pm$0.01\\
H II       &-19.26$\pm$0.05&-19.59$\pm$0.02& &2.38$\pm$0.01 &2.34$\pm$0.01\\
\enddata 
\tablenotetext{a}{The Concentration index ($C$) is defined by the ratio
  $C = r90/r50$, where $r90$ and $r50$ correspond to the radii at
  which the integrated fluxes (in $r$-band) are equal to 90\% and 50\%
  of the Petrosian flux respectively.}
\tablecomments{Central values are medians of parameters listed in the
  header.  Errors are the standard deviation of the median 
($ = 1.253 \times
  \sigma_{\rm mean}$, assuming Gaussian distributions; Lupton 1993).}
\end{deluxetable}


\begin{deluxetable}{lccccc}
\tablecolumns{10} \tablewidth{0pt} 
\tablecaption{ Radio Properties
\label{radiodata}}
\tablehead{ 
\colhead{} & 
\multicolumn{2}{c}{$N_{obj}$ (Fraction)\tablenotemark{a}} &
\colhead{} & 
\multicolumn{2}{c}{$\nu L_{\nu}$ (1.4GHz)\tablenotemark{b}}\\ 
\cline{2-3} \cline{5-6}\\ 
\colhead{Sample} &
\colhead{void} & \colhead{wall} & \colhead{} & \colhead{void} &
\colhead{wall} } 
\startdata 
Seyfert    & 2 (15\%)  &38 (20\%) & &0.9$\pm$0.1&4.1$\pm$1.8\\ 
LINER      & 0 (0\%)   &35 (7\%)  & & \nodata   &6.9$\pm$1.5\\ 
Transition & 5 (9\%)   &74 (9\%)  & &1.6$\pm$0.7&3.7$\pm$1.2\\ 
H II       & 9 (2.7\%) &91 (3.4\%)& &0.9$\pm$0.2&1.6$\pm$0.1\\ 
\enddata 
\tablenotetext{a} {counts of sources with FIRST detections at the flux
density threshold of 1mJy at 1.4 GHz.}
\tablenotetext{b}{in units of $10^{38}$ erg s$^{-1}$.  
Central values are medians and errors are the standard deviation of 
the median.}  
\end{deluxetable}


\begin{deluxetable}{lccccccrc}
\tablecolumns{9} \tablewidth{0pt} 
\tablecaption{Nearest Neighbor (nn)
Statistics
\label{nn}}
\tablehead{
\colhead{} &
\multicolumn{2}{c}{$d_{\rm 3nn}$} &
\colhead{} &
\multicolumn{2}{c}{$d_{\rm 1nn}$} &
\colhead{} &
\multicolumn{2}{c}{$d_{\rm 1nn}$, whole sample}\\
\cline{2-3} \cline{5-6} \cline{8-9}\\
\colhead{Sample} &
\colhead{void} &
\colhead{wall} &
\colhead{} &
\colhead{void} &
\colhead{wall} &
\colhead{} &
\colhead{void} &
\colhead{wall} }
\startdata
Seyfert    &8.4 $\pm$ 0.7&~3.1 $\pm$ 0.1~~~& &5.8 $\pm$ 0.9&~1.6 $\pm$ 0.1~~& &2.8 $\pm$ 0.8&~1.1 $\pm$ 0.1~~\\
LINER      &8.6 $\pm$ 0.4&~2.7 $\pm$ 0.1~~~& &5.9 $\pm$ 0.6&1.34 $\pm$ 0.05& &2.1 $\pm$ 0.7&0.92 $\pm$ 0.05\\
Transition &8.3 $\pm$ 0.4&~3.0 $\pm$ 0.1~~~& &4.8 $\pm$ 0.4&1.77 $\pm$ 0.05& &2.8 $\pm$ 0.4&1.02 $\pm$ 0.04\\
H II       &8.5 $\pm$ 0.1&3.36 $\pm$ 0.04  & &4.4 $\pm$ 0.2&1.85 $\pm$ 0.03& &2.0 $\pm$ 0.2&1.07 $\pm$ 0.03\\
\enddata 
\tablecomments{Central values are medians of distances to the
  3rd and 1st nearest neighbor, in units of $h^{-1}$Mpc.  Errors are
  the standard deviation of the median.
  The first two columns indicate measurements for objects in the $M_r < -19.5$
  volume-limited sample only.  The whole sample includes all objects with 
  $14.5 < m_r < 17.7$, and thus adds intrinsically faint sources.}
\end{deluxetable}

\begin{thebibliography}{}
\bibitem[e.g., Baldwin, Phillips, \& Terlevich (1981)]{bpt81} Baldwin,
J. A.  Phillips, M. M, \& Terlevich, R., 1981, \pasp, 93, 5
\bibitem[e.g., Balogh et al. (2004)]{bal04} Balogh, M., et al., 2004,
  \mnras, 348, 1355
\bibitem[Barth \& Shields (2000)]{bar00} Barth, A., \& Shields, C. J.,
2000, \pasp, 112, 753
\bibitem[FIRST; Becker,White \& Helfand (1995)]{bec95} Becker, R.  H.,
  White, R. L., \& Helfand, David J., 1995, \apj, 450, 559
\bibitem[Begelman \& Nath (2005)]{beg05} Begelman, M. C., \& Nath,
  B.B.,  2005, \mnras, 361, 1387
\bibitem[Benson et al. (2003)]{ben03} Benson, A. J., Hoyle, F., Torres,
F \& Vogeley, M. S., 2003, MNRAS, 340, 160
\bibitem[Best (2004)]{bes04} Best, P. N., 2004, \mnras, 351, 70
\bibitem[Blanton et al. (2003)]{bla03}Blanton, M. R., Lupton, R. H.,
Maley, F. M., Young, N., Zehavi, I., \& Loveday, J. 2003, \aj, 125,
2276
\bibitem[Blanton et al. (2005)]{bla05}Blanton, M. R., et al. 2005,
  \aj, 129, 2562
\bibitem[Brinchmann et al. (2004)]{bri04} Brinchmann, J., Charlot, S.,
Heckman, T. M., Kauffmann, G., Tremonti, C., \& White, S.D.M., 2004,
astro-ph/0406220 
\bibitem[Byrd et al. (1986)]{byr86} Byrd, G. G., Sundelius, B., \&
Valtonen, M., 1986, A\&A, 171, 16
\bibitem[Christlein \& Zabludoff (2005)]{chr05} Christlein, D., \&
Zabludoff, A. I., 2005, \apj, 621, 201
\bibitem[Constantin \& Vogeley (2006)]{con06a} Constantin, A., and
  Vogeley, M.S., 2006, \apj, 650, 727
\bibitem[Constantin et al. (2007)]{con07} Constantin, A., et al.,
  2007, \apj, submitted.
\bibitem[Croton et al. (2006)]{cro06} Croton et al., 2006, \mnras,
  365, 11
\bibitem[Cruzen et al. (2002)]{cru02} Cruzen et al., 2002, \apj, 123,
  142
\bibitem[Dahari (1984)]{dah84} Dahari, O., 1984, \aj, 89, 966
\bibitem[de Robertis et al. (1998a)]{derob98a} de Robertis, M. M.,
  Hayhoe, K., \& Yee, H. K. C., 1998, \apjs, 115, 163
\bibitem[de Robertis et al. (1998b)]{derob98b} de Robertis, M. M.,
  Yee, H. K. C., \& Hayhoe, K., 1998, \apj, 496, 93
\bibitem[e.g., DiMatteo et al. (2005)]{dimat05} DiMatteo et al., 2005,
  Nature, 433, 604
\bibitem[Dopita \& Sutherland (1995)]{dop95} Dopita, M. A. \&
Sutherland, R. S. 1995, ApJ, 455, 468
\bibitem[Doyle \& Drinkwater (2006)]{doy06} Doyle, M. T., \&
  Drinkwater, M. J., 2006, \mnras, 372, 977
\bibitem[Einasto et al. (1980)]{ein80} Einasto, J., Joeveer, M., Saar,
  E., 1980, \mnras, 193, 353
\bibitem[Eisenstein et al. (2001)]{eis01} Eisenstein, D. J., et al. 2001,
AJ, 122, 2267
\bibitem[Ferrarese \& Merrit (2000)]{fer00} Ferrarese, L. \&
Merritt, D., 2000, ApJ, 539, 9
\bibitem[Filippenko \& Terlevich (1992)]{fil92} Filippenko, A. V., \&
Terlevich, R., 1992, \apj, 397, 79L
\bibitem[Fuentes-Williams \& Stocke (1988)]{fue88} Fuentes-Williams,
  T., \& Stocke, J. T., 1988, \aj, 96, 1235
\bibitem[Fukugita et al. (1996)]{fuk96} Fukugita, M., Ichikawa, T.,
Gunn, J. E., Doi, M., Shimasaku, K., \& Schneider, D. P., 1996, \aj,
111, 1748
\bibitem[G\'{o}mez et al. (2003)]{gom03} G\'{o}mez, P. L., et al. 2003, \apj,
  584, 210
\bibitem[Gebhardt et al. (2000)]{geb00} Gebhardt, K., et al., 2000,
\apj, 539, 13
\bibitem[Goldberg et al. (2005)]{gol05} Goldberg D. M., Jones, Timothy
D., Hoyle, F., Rojas, R. R., Vogeley, M. S., \& Blanton,
M. R., 2005, \apj, 621, 643
\bibitem[Greene et al. (2004)]{gre04} Greene, J.E., Lim, J., \& Ho,
  P. T. P., 2004, \apjs, 153, 93 
\bibitem[Gunn et al. (1998)]{gun98}Gunn, J. E., et al. 1998, \apj,
  116, 3040
\bibitem[Heckman (1980)]{hec80} Heckman T. M., 1980, A\&A, 87, 152
\bibitem[Heckman et al. (2004)]{hec04} Heckman, T. M., Kauffmann, G.,
Brinchmann, J., Charlot, S., Tremonti, C., White, S.
D. M., 2004, \apj, 613, 109
\bibitem[Hernquist \& Mihos (1995)]{her95} Hernquist, L., \& Mihos,
  J. C., 1995, \apj, 448, 41
\bibitem[Ho et al. (1997a)]{ho97} Ho, L. C., Filippenko, A. V., \&
Sargent, W. L. W., 1997, \apjs, 112, 315
\bibitem[e.g., Ho et al. (1997b)]{ho97morph} Ho, L. C., Filippenko,
A. V., \& Sargent, W. L. W., 1997, \apjs, 487, 568
\bibitem[Hogg et al. (2001)]{hog02} Hogg, D. W., Finkbeiner, D. P.,
Schlegel, D. J., \& Gunn, J. E. 2001, \aj, 122, 2129
\bibitem[Hopkins et al. (2005)]{hop05} Hopkins, P. F., Hernquist, L.,
Cox, T. J., Di Matteo, T., Martini, P., Robertson, B., \& Springel, V.,
2005, \apj, 630, 705
\bibitem[e.g., Hopkins et al. (2006)]{hop06} Hopkins, P. F.,
Hernquist, L., Cox, T. J., Di Matteo, T., Robertson, B., \& Springel,
V., 2006, \apj, 163, 1
\bibitem[Hoyle \& Vogeley (2002)]{HV02} Hoyle, F. \& Vogeley, M. S.,
2002, ApJ, 566, 641
\bibitem[Hoyle \& Vogeley (2004)]{HV04} Hoyle, F. \& Vogeley, M. S.,
2004, \apj, 607, 751
\bibitem[Hoyle et al. (2005)]{hoy05} Hoyle, F. Rojas, R. R., Vogeley,
M. S. \& Brinkmann, J., 2005, \apj, 620, 618
\bibitem[Kauffmann \& Haehnelt (2000)]{kau00} Kauffmann, G., \&
  Haehnelt, M., 2000, 311, 576
\bibitem[Kauffmann et al. (2003a)]{kau03b} Kauffmann, G., et al., 2003,
  \mnras, 341, 33
\bibitem[Kauffmann et al. (2003b)]{kau03c} Kauffmann, G., et al., 2003,
  \mnras, 341, 54
\bibitem[Kauffmann et al. (2003c)]{kau03} Kauffmann, G., et al., 2003,
  \mnras, 346, 1055
\bibitem[Kauffmann et al. (2004)]{kau04} Kauffmann, G., et al., 2004,
  \mnras, 353, 713
\bibitem[Kewley et al. (2001)]{kew01} Kewley, L., Dopita, M. A.,
Sutherland, R. S., Heisler, C. A., \& Trevena, J., 2001, \apj, 556, 121
\bibitem[Kewley et al. (2006)]{kew06} Kewley, L.J., Groves, B.,
  Kauffmann, G., \& Heckman, T., 2006, \mnras, 372, 961
\bibitem[Kirshner et al. (1981)]{kir81} Kirshner, R. P., Oemler, A.,
Jr., Schechter, P. L., \& Shectman, S. A., 1981, \apj, 248, 57
\bibitem[Laine et al. (2002)]{lai02} Laine, S., Shlosman, I., Knapen,
J. H., Peletier, R. F., 2002, \apj, 567, 97
\bibitem[Laurikainen \& Salo (1995)]{lau95} Laurikainen, E., \& Salo,
  H., 1995, A\&A, 293, 683
\bibitem[Lewis et al. (2002)]{lew02} Lewis, I. et al., 2002, \mnras,
334, 673
\bibitem[e.g., Lupton (1993)]{lup93} Lupton, R.H., 1993, {\it
  Statistics in Theory and Practice}, Princeton: Princeton University
  Press
\bibitem[Lupton et al. (2002)]{lup02} Lupton, R.H., Ivezic, Z., Gunn,
J. E., Knapp, G., Strauss, M. A., Yasuda, N., 2002, Proc. SPIE, 4836, 350
\bibitem[Magorrian et al. (1998)]{mag98} Magorrian, J., et al., 1998,
  \aj, 115, 2285
\bibitem[e.g., Malkan et al. (1998)]{mal98} Malkan, M. A., Gorjian,
  V., \& Tam, R., 1998, \apjs, 117, 25
\bibitem[Marconi \& Hunt (2003)]{mar03} Marconi, A., \& Hunt, L. K.,
2003, \apj, 589, 21
\bibitem[Miller et al. (2003)]{mil03} Miller, C. J., Nichol, R. C.,
G\'omez, P. L., Hopkins, A. M., \& Bernardi, M., 2003, \apj, 597, 142
\bibitem[e.g., Mulcheay \& Regan (1997)]{mul97} Mulchaey, John S., \&
Regan, Michael W., 1997, \apj, 482, 135
\bibitem[Narayan \& Yi (1994)]{nar94} Narayan, R., Yi, I., 1994, \apj,
  428, 13
\bibitem[Osterbrock (1989)]{ost89} Osterbrock, D. E., 1989, {\it
Astrophysics of Gaseous Nebulae and Active Galactic Nuclei},
University Science Books
\bibitem[Park et al. (2007)]{park07} Park, C., Choi, Y., Vogeley,
M. S., Gott, J. Richard, III, \& Blanton, Michael R., 2007, \apj, 658,
898
\bibitem[Pier et al. (2003)]{pie03} Pier, J. R., Munn, J. A., Hindsley,
R. B., Hennessy, G. S., Kent, S . M., Lupton, R. H., \& Ivezi\'{c},
\v{Z}. 2003, \aj, 125, 1559
\bibitem[Quataert (2003)]{qua03} Quataert, E., 2003, Astron.
  Nachr., Supp. Issue 1, in Proceedings of the Galactic
  Center Workshop 2002 - {\it The central 300 parsecs of the Milky Way},
  p.435-443 
\bibitem[Rafanelli et al. (1995)]{raf95} Rafanelli, P., Violato, M.,
  \& Baruffolo, A., 1995, \aj, 109, 1546
\bibitem[e.g., Rees et al. (1982)]{ree82} Rees, M. J., Begelman,
  M. C., Blandford, R. D., Phinney, E. S., 1982, Nature, 295, 17
\bibitem[Rojas et al. (2004)]{rojasphoto} Rojas, R. R., Vogeley,
M. S., Hoyle, F. \& Brinkmann, J., 2004, \apj, 617, 50
\bibitem[Rojas et al. (2005)]{rojasspectro} Rojas, R. R., Vogeley,
M. S., Hoyle, F. \& Brinkmann, J., 2005, \apj, 624, 571
\bibitem[e.g., Schawinski et al. (2006)]{sch06} Schawinski, K., et
  al., 2006, Nature, 442, 888
\bibitem[Schawinski et al. (2007)]{sch07} Schawinski, K., et
  al., 2007, \mnras, in press (astro-ph/0709.3015)
\bibitem[Schmitt (2001)]{sch01} Schmitt, H. R., 2001, \aj,  122, 2243
\bibitem[Schmitt (2004)]{sch04} Schmitt, H. R., 2004, in {\it The
Interplay among Black Holes, Stars and ISM in Galactic Nuclei},
Cambridge Univ. Press, Proceedings of IAU Symposium, No. 222,
eds. T. Storchi-Bergmann, L.C. Ho, and Henrique R. Schmitt, p. 395.
\bibitem[e.g., Silk \& Rees (1998)]{silk98} Silk, J., \& Rees, M.J.,
  1998, A\&A, 331, 1
\bibitem[Springel et al. (2005)]{spr05} Springel, V., Di Matteo, T.,
  \& Hernquist, L., 2005, \mnras, 361, 776
\bibitem[Shields (1992)]{shi92} Shields, J. C., 1992, \apj, 399, 27L
\bibitem[Shields et al. (2006)]{shi06} Shields, J. C., Rix, H.-W.,
Sarzi, M., Barth, A. J., Filippenko, A. V., Ho, L. C., McIntosh,
D. H., Rudnick, G., Sargent, W. L. W., 2007, \apj, 654, 125
\bibitem[Smith et al. (2002)]{smi02} Smith, J. A., et al. 2002, AJ, 123,
2121
\bibitem[Strauss et al. (2002)]{str02} Strauss, M., et al., 2002, \aj,
  124, 1810
\bibitem[Taniguchi, Shioya, \& Murayama (2000)]{tan00} Taniguchi, Y.,
Shioya, Y., \& Murayama, T., 2000, \aj, 120, 1265
\bibitem[Tremonti et al. (2004)]{tre04} Tremonti, C.A., et al., 2004,
  \apj, 613, 898
\bibitem[Veilleux \& Osterbrock (1987)]{vei87} Veilleux, S., \&
Osterbrock, D. E., 1987, \apjs, 63, 295
\bibitem[e.g., Wild et al. (2007)]{wild07} Wild, V., Kauffmann, G.,
Heckman, T., Charlot, S., Lemson, G., Brinchmann, J., Reichard, T., \&
Pasquali, A., 2007, \mnras, submitted (astro-ph/0706.3113)
\bibitem[York et al. (2000)]{yor00} York, D. G., et al., 2000, \aj,
120,1579


\end{thebibliography}
\end{document}